\pdfoutput=1
\documentclass{JINST}

\usepackage{amsmath}
\usepackage{rotating}
\usepackage{url} 
\usepackage{booktabs}

\newcommand{\lighttoprule}{\toprule[\lightrulewidth]}

\newcommand{\kforty}         {\ensuremath{^{40}\text{K}}}
\newcommand{\mnfiftyfour}    {\ensuremath{^{54}\text{Mn}}}
\newcommand{\cofiftysix}     {\ensuremath{^{56}\text{Co}}}
\newcommand{\cofiftyeight}   {\ensuremath{^{58}\text{Co}}}
\newcommand{\fefiftynine}    {\ensuremath{^{59}\text{Fe}}}
\newcommand{\cosixty}        {\ensuremath{^{60}\text{Co}}}
\newcommand{\znsixtyfive}    {\ensuremath{^{65}\text{Zn}}}

\newcommand{\xeonethirtysix} {\ensuremath{^{136}\text{Xe}}}

\newcommand{\bitwoten}       {\ensuremath{^{210}\text{Bi}}}
\newcommand{\potwoten}       {\ensuremath{^{210}\text{Po}}}
\newcommand{\pbtwoten}       {\ensuremath{^{210}\text{Pb}}}
\newcommand{\rntwotwentytwo} {\ensuremath{^{222}\text{Rn}}}
\newcommand{\thtwothirtytwo} {\ensuremath{^{232}\text{Th}}}
\newcommand{\utwothirtyeight}{\ensuremath{^{238}\text{U}}}

\newcommand{\gr}        {\ensuremath{\gamma}\text{-ray}}
\newcommand{\qbb}       {\ensuremath{Q_{\beta\beta}}}
\newcommand{\twonubb}   {\ensuremath{2\nu\beta\beta}}
\newcommand{\zeronubb}  {\ensuremath{0\nu\beta\beta}}

\renewcommand{\deg}{\ensuremath{^\circ}}

\newcommand{\znbb}{\textnormal{0}\nu\beta\beta}
\newcommand{\tnbb}{\textnormal{2}\nu\beta\beta}

\newcommand{\xe}{^{136}\textnormal{Xe}}

\title{The EXO-200 detector, part I: Detector design and construction}

\author{M.~Auger$^a$, D.J.~Auty$^b$, P.S.~Barbeau$^c$, L.~Bartoszek$^c$\footnote{Also Bartoszek Engineering, Aurora, IL, USA.}, E.~Baussan$^a$\footnote{Now at the Dept. Recherches Subatomiques, Institut Pluridisciplinaire H. Curien, Strasbourg, France}, E.~Beauchamp$^d$, C.~Benitez-Medina$^e$, M.~Breidenbach$^f$, D.~Chauhan$^d$, B.~Cleveland$^d$\footnote{Also SNOLAB, Sudbury ON, Canada}, R.~Conley$^f$, J.~Cook$^g$, S.~Cook$^e$, A.~Coppens$^h$, W.~Craddock$^f$, T.~Daniels$^g$, C.G.~Davis$^i$, J.~Davis$^c$, R.~deVoe$^c$, A.~Dobi$^i$, M.J.~Dolinski$^c$\thanks{Corresponding author.}~, M.~Dunford$^h$, W.~Fairbank Jr.$^e$, J.~Farine$^d$, P.~Fierlinger$^j$, D.~Franco$^a$, G.~Giroux$^a$, R.~Gornea$^a$,K.~Graham$^h$, G.~Gratta$^c$, C.~Hagemann$^h$, C.~Hall$^i$, K.~Hall$^e$, C.~Hargrove$^h$, S.~Herrin$^f$, J.~Hodgson$^f$, M.~Hughes$^b$, A.~Karelin$^k$, L.J.~Kaufman$^l$, J.~Kirk$^c$, A.~Kuchenkov$^k$, K.S.~Kumar$^g$, D.S.~Leonard$^m$, F.~Leonard$^h$, F.~LePort$^c$\footnote{Now at TESLA Motors, Palo Alto CA, USA}, D.~Mackay$^f$\footnote{Now at KLA-Tencor, Milpitas CA, USA}, R.~MacLellan$^b$, M.~Marino$^j$, K.~Merkle$^c$, B.~Mong$^d$, M.~Montero D\'{i}ez$^c$, A.R.~M\"{u}ller$^c$\footnote{Now at Kayser-Threde, Munich, Germany},R.~Neilson$^c$\footnote{Now at Department of Physics, University of Chicago, Chicago IL, USA}, A.~Odian$^f$, K.~O'Sullivan$^c$, C.~Ouellet$^h$, A.~Piepke$^b$, A.~Pocar$^g$, C.Y.~Prescott$^f$, K.~Pushkin$^b$, A.~Rivas$^c$, E.~Rollin$^h$, P.C.~Rowson$^f$, A.~Sabourov$^c$, D.~Sinclair$^h$\footnote{Also TRIUMF, Vancouver BC, Canada}, K.~Skarpaas$^f$, S.~Slutsky$^i$, V.~Stekhanov$^k$, V.~Strickland$^h$\footnotemark[9], M.~Swift$^f$, D.~Tosi$^c$, K.~Twelker$^c$, J.-L.~Vuilleumier$^a$, J.-M.~Vuilleumier$^a$, T.~Walton$^e$, M.~Weber$^a$, U.~Wichoski$^d$, J.~Wodin$^f$, J.D~Wright$^g$, L.~Yang$^f$, Y.-R.~Yen$^i$\\
\llap{$^a$}LHEP, Albert Einstein Center, University of Bern, Bern, Switzerland\\
\llap{$^b$}Department of Physics and Astronomy, University of Alabama, Tuscaloosa AL, USA\\
\llap{$^c$}Physics Department, Stanford University, Stanford CA, USA\\
\llap{$^d$}Physics Department, Laurentian University, Sudbury ON, Canada\\
\llap{$^e$}Physics Department, Colorado State University, Fort Collins CO, USA\\
\llap{$^f$}SLAC National Accelerator Laboratory, Stanford CA, USA,\\
\llap{$^g$}Physics Department, University of Massachusetts, Amherst MA, USA\\
\llap{$^h$}Physics Department, Carleton University, Ottawa ON, Canada\\
\llap{$^i$}Physics Department, University of Maryland, College Park MD, USA\\
\llap{$^j$}Excellence Cluster `Universe', Technische Universit\"{a}t M\"{u}nchen, Garching, Germany\\
\llap{$^k$}Institute for Theoretical and Experimental Physics, Moscow, Russia\\
\llap{$^l$}Physics Department and CEEM, Indiana University, Bloomington IN, USA\\
\llap{$^m$}Physics Department, University of Seoul, Seoul, Korea\\
E-mail: \email{dolinski@stanford.edu}}

\abstract{EXO-200 is an experiment designed to search for double beta decay of $^{136}$Xe with a single-phase, liquid xenon detector.  It uses an active mass of 110~kg of xenon enriched to 80.6\% in the isotope 136 in an ultra-low background time projection chamber capable of simultaneous detection of ionization and scintillation.  This paper describes the EXO-200 detector with particular attention to the most innovative aspects of the design that revolve around the reduction of backgrounds, the efficient use of the expensive isotopically enriched xenon, and the optimization of the energy resolution in a relatively large volume.}


\begin{document}

\section{Introduction}
Double beta decay is the dominant decay mode for certain even-even nuclei for which beta decay is energetically forbidden or suppressed by a large change in angular momentum.  The process is rare due to its second-order nature and can only be directly observed in specially built, low background setups.

Two-neutrino double beta decay ($\tnbb$) is allowed by the Standard Model and has been observed in the laboratory for several isotopes.  Its half-life is typically $\sim 10^{20}$~years as first calculated in~\cite{MariaGoppertMayer}.  Zero-neutrino double beta decay ($\znbb$), on the other hand, is a lepton-number violating decay that can occur only if neutrinos are massive Majorana particles~\cite{Boehm-Vogel, Schechter:1981bd}.   Its detection is considered the most sensitive probe of the neutrino mass scale and the question of whether 2-component Majorana particles, first discussed in~\cite{Majorana, Racah}, exist in nature.  The recent discovery of neutrino oscillations~\cite{oscillations1, oscillations2} has firmly established the existence of finite neutrino masses, yet without providing clues either to their nature (Dirac or Majorana) or to their absolute values.   This provides the motivation for an extensive program searching for the $\znbb$ decay with ultimate sensitivities to Majorana masses close to or below 10~meV.

There are several experiments designed to search for $\znbb$ in a number of different candidate isotopes~\cite{Avignone:2007fu}.  There is also a claim for the observation of $\znbb$ in $^{76}$Ge~\cite{HVKK}.  One of the isotopes of interest for $\znbb$ searches is $\xe$.  $\xe$ is an attractive candidate nuclide because it has a high Q-value located in a region relatively free from naturally occurring radioactive backgrounds.  In addition, isotopic enrichment of Xe is simpler than in other cases.  $\xe$ can also form the detection medium in a noble liquid or gas time projection chamber (TPC), either of which can be operated with continuous inline purification of the Xe.

The Enriched Xenon Observatory (EXO) Collaboration is planning a series of experiments to search for $\znbb$ of $\xe$ with progressively higher sensitivity.    Within this program, EXO-200 is a 100~kg-scale experiment designed to achieve a sensitivity close to 100~meV for Majorana neutrino masses.   EXO-200 started low-background data taking in May 2011 and has recently reported the first observation of $\tnbb$ in $\xe$~\cite{Ackerman}.  EXO-200 was also designed to serve as the prototype for a multi-ton detector with Majorana mass sensitivity below 10~meV.  

This paper reviews the design and construction of the EXO-200 detector.   The cryogenics, controls, vacuum, and other infrastructure, allowing the TPC to operate within a narrow temperature and pressure window, will be discussed in a separate paper.   A third paper will describe the performance of the detector system.

\section{EXO-200 detector design and scientific reach}

\subsection{Overview of the EXO-200 detector}

EXO-200 was designed to be a state of the art double beta decay experiment and, at the same time, a technology test bed for a future, larger detector.   In order to take advantage of event topology, xenon self-shielding, and the possibility of purifying a noble element before and during its use, EXO-200 uses the xenon as both source and detector in a homogeneous, liquid phase TPC~\cite{TPC}.   At the operating temperature (167~K) and pressure (147~kPa) the liquid xenon (LXe) has a density of 3.0~g/cm$^3$~\cite{density}. The xenon for EXO-200 is enriched to 80.6\% in the isotope $\xe$.

\begin{figure}
	\centering
	\includegraphics[width=5in]{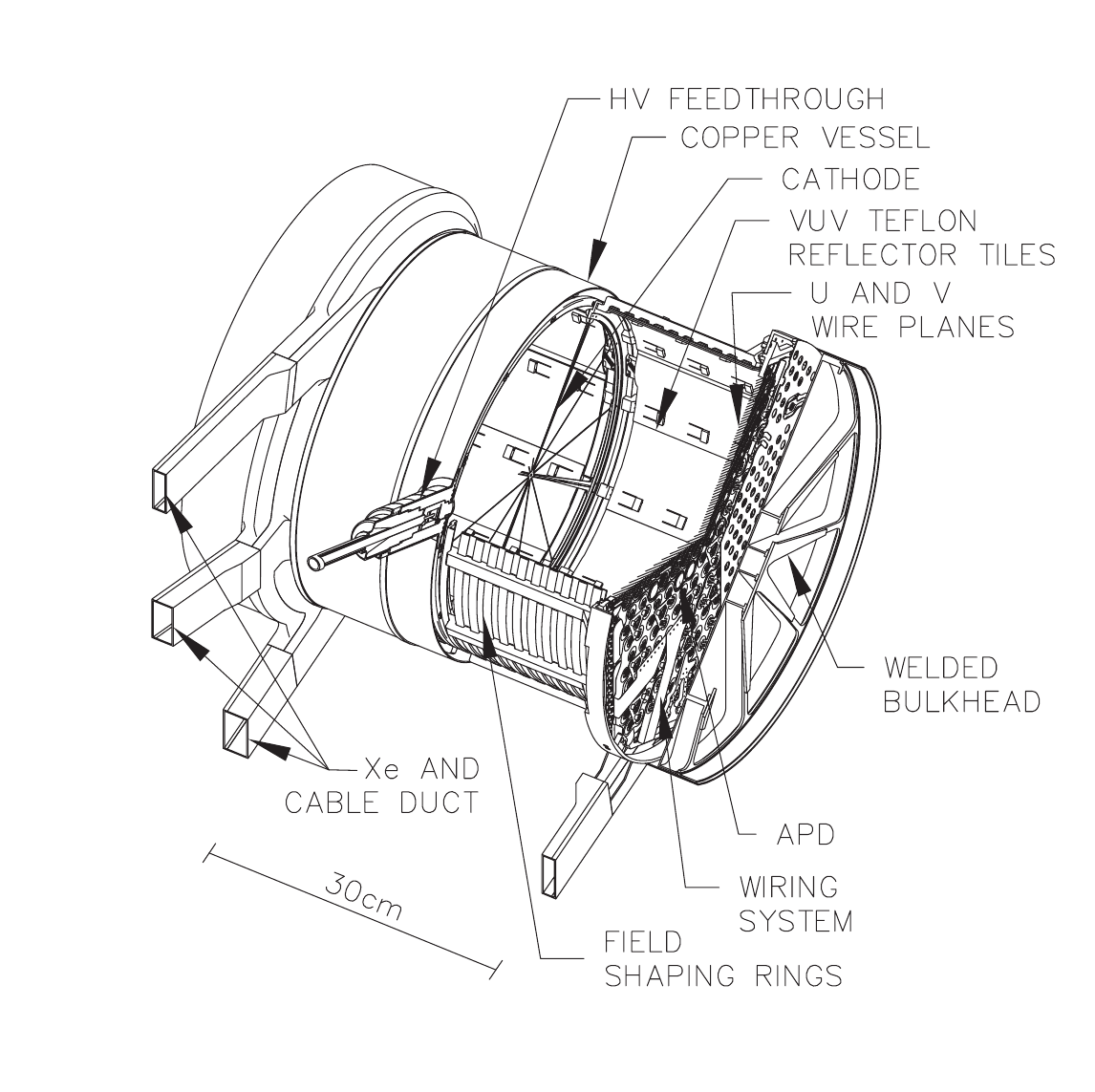}
	\caption{Cutaway view of the EXO-200 TPC with the main components identified.}
	\label{fig:TPC}
\end{figure}

In order to minimize the surface-to-volume ratio while maintaining a practical geometry, the detector is a double TPC, having the shape of a square cylinder with a cathode grid held at negative high voltage at the mid plane.   The signal readout is performed at each base of the cylinder, near ground potential.   Of the 200~kg of enriched xenon available, 175~kg are in liquid phase, and 110~kg are in the active volume of the detector.  A cutaway view of the TPC is shown in Figure~\ref{fig:TPC}.

Two considerations were central in designing the detector: the need for good energy resolution at the double beta decay decay $Q$-value of 2457.8~keV~\cite{Q-value}, and the requirement to achieve exceedingly low backgrounds.   Early R\&D performed by the EXO collaboration~\cite{Conti_etal} showed that the energy resolution in LXe can be substantially improved by using an appropriate linear combination of ionization and scintillation as the energy estimator.  This technique was subsequently used in other contexts~\cite{Aprile3}.   In EXO-200 both the ionization and the scintillation signals are recorded.  Charge is collected at each end of the TPC by wire planes, held at virtual ground, while the 178~nm-wavelength scintillation light is collected by two arrays of large area avalanche photodiodes (LAAPDs)~\cite{Neilson:2009kf}, one behind each of the two charge collection planes (``U'' wires).  The decision to use LAAPDs instead of photomultiplier tubes, unusual among large LXe detectors, is important because they combine high quantum efficiency for the scintillation light with ultra-low levels of radioactivity.    The drawback of this choice is that the noise in the LAAPDs, while modest at 170~K, limits the scintillation threshold of the detector.  The 468 LAAPDs are used as bare dies submerged in the LXe, avoiding substrates that are often the main source of radioactivity and ensuring that the devices are not mechanically stressed at cryogenic temperatures.  A second wire plane (``V'' wires), positioned in front of the charge collection plane and oriented at 60$^{\circ}$ from it, is biased to ensure full electron transparency and is used to inductively record a second coordinate for each ionization cluster.  Three-dimensional position sensitivity is achieved by using the difference in the arrival time between the ionization and scintillation signals to calculate the electron drift time.

Because the rate of $\znbb$ is so small, it is important to minimize backgrounds that can deposit energy near the $\xe$ $Q$-value. The attenuation length in LXe of 2.5~MeV \gr s is quite large ($\sim9$~cm)~\cite{attenuation}, so xenon self-shielding is not particularly effective, and the materials near the active Xe volume have to be intrinsically clean (this is in contrast to detectors built for dark matter searches that are optimized for lower energies where self-shielding is substantially more effective).   Very low backgrounds are achieved by selecting ultra-low radioactivity construction materials, minimizing the masses of passive components through careful design, specially cleaning and storing components before assembly,  and building the detector in progressively clean, shielded layers.  The TPC was built using primarily copper and bronze for conductors and acrylic, PTFE and polyimide for dielectrics.  Some of the specific materials used are reported in~\cite{Background_NIM}.   All TPC materials were degreased and etched to remove surface contamination.

The LXe container (``vessel'') was machined from selected copper stock and assembled using electron-beam welding.   In order to reduce the activity nearest to the LXe, the copper vessel was built with a 1.37~mm thickness that is designed to reliably support only a 35~kPa pressure differential (in either direction).  The copper vessel is designed to closely envelop the active Xe volume, flaring out at the ends to contain the frames of the wire planes, the LAAPDs, and the wiring.   The vessel is welded shut at the two ends with a TIG field weld to minimize the use of materials and avoid sealing problems. An elaborate control system (to be described, along with the cryostat, in a future paper) ensures that the pressure inside the TPC vessel tracks the pressure outside to within a tight tolerance ($\sim\pm$4~kPa) during detector pump down, purge, cooling, liquid fill, and normal operations.   

\begin{figure}
	\centering
	\includegraphics[width=6in]{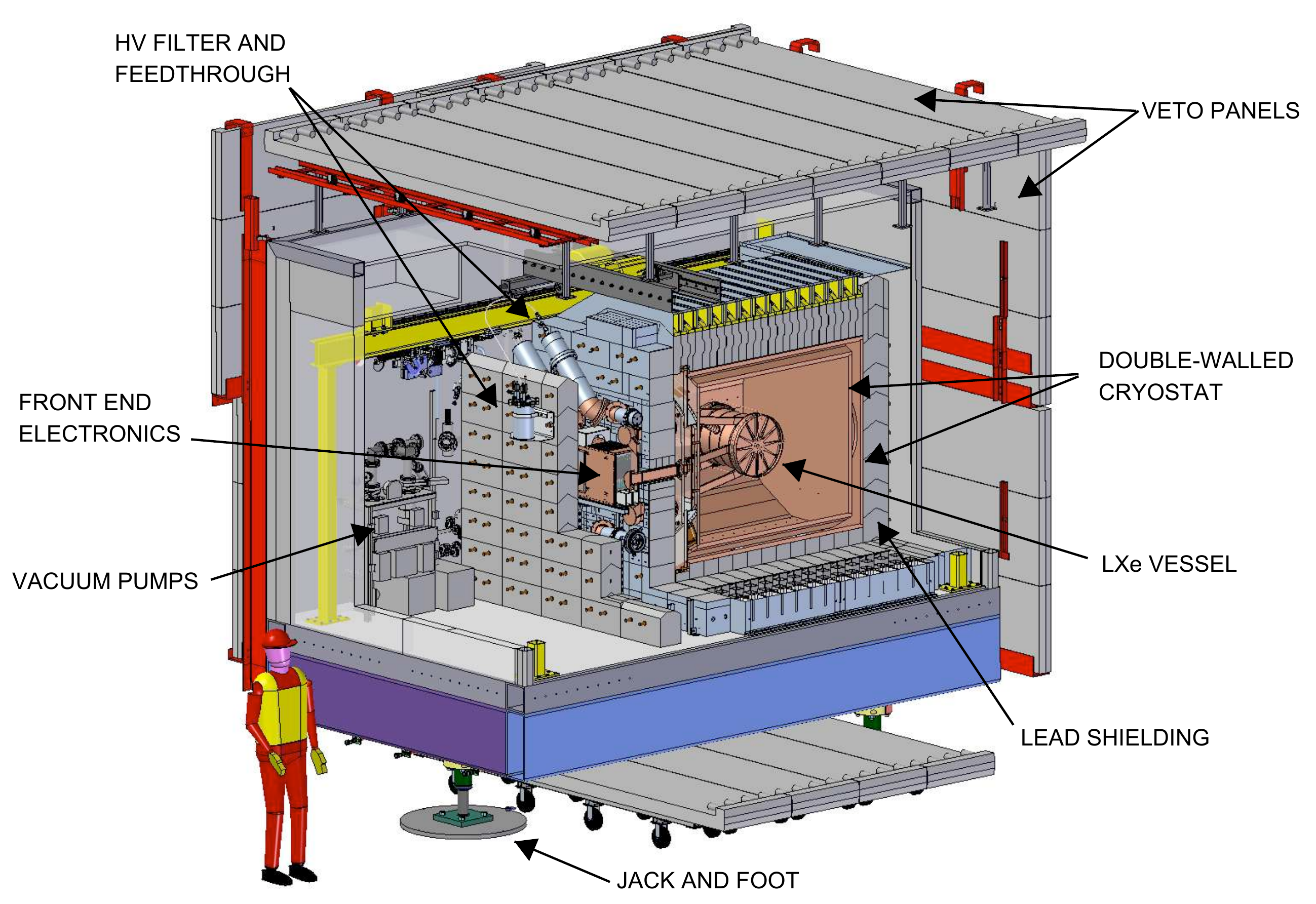}
	\caption{Cutaway view of the EXO-200 setup, with the primary subassemblies identified.}
	\label{fig:cleanroom}
\end{figure}

Figure~\ref{fig:cleanroom} shows a cutaway view of the entire detector complex, including a double-walled copper cryostat, lead shielding, and muon veto.  The cryostat is a twelve-sided, double walled, vacuum insulated copper vessel, made from specially selected low background copper. The copper was freshly produced in a dedicate low background run for EXO by the Norddeutsche Affinerie (now Aurubis, Germany~\cite{aurubis}) to minimize its exposure to the cosmic radiation. The cryostat contains a total of 5901~kg of copper.  All these components, with the exception of the muon veto, are located inside a class 1000 clean room. The inner cryostat contains an ultra-clean, dense ($\rho=1.8$~g/cm$^2$ at 170~K) fluid (HFE-7000~\cite{HFE}), providing both shielding and thermal uniformity. The fluid also transfers the pressure load to the 25~mm thick inner cryostat copper vessel that is designed to tolerate absolute implosive (explosive) loads $>$100~kPa ($>$300~kPa).

The outermost  shielding layer, outside the outer vessel of the cryostat, consists of 25~cm of lead.  The low-noise front end electronics are located outside of the lead shielding and are connected to the detector through thin polyimide cables.   This choice trades some increased noise for the simplicity and accessibility of room temperature, conventional construction electronics.

A cosmic-ray veto counter made of plastic scintillators surrounds the cleanroom housing the rest of the detector.   EXO-200 is located at a depth of 1585~m water equivalent~\cite{muonflux} in the Waste Isolation Pilot Plant (WIPP) near Carlsbad, New Mexico (32$^{\circ}$22'30''N 103$^{\circ}$47'34''W).   

\subsection{Design sensitivity and estimated backgrounds}

While the measured performance of EXO-200 utilizing substantial low-background and calibration data sets will be the subject of a future paper, here we provide sensitivity figures assuming design parameters for the detector performance and background.   Initial data taking roughly confirms the validity of such parameters. Using the expected energy resolution of ${\sigma_E}/E = 1.6\%$ at the $^{136}$Xe end point, EXO-200 was designed to reach a sensitivity of $T^{0\nu\beta\beta}_{1/2}=6.4\times 10^{25}$~yr (90\%~C.L.) in two years of live time, should the $\znbb$ be beyond reach. $\znbb$ is defined by a $\pm 2\sigma$ window around the end-point and 40 background events are expected to accumulate in such a window in two years.  This estimate was made using a fiducial mass of 140~kg (200~kg with 70\% efficiency), while the final detector design has 110~kg of active Xe, requiring a longer time to reach the same sensitivity.  The $T_{1/2}$ limit above corresponds to a 90\%~C.L. Majorana mass sensitivity of 109~meV (135~meV) using the QRPA~\cite{qrpa} (NSM~\cite{nsm}) matrix element calculation.

EXO-200 was designed to have sufficient sensitivity to confirm or reject the claim of $\znbb$ discovery by~\cite{HVKK}.   Using the value $T^{0\nu\beta\beta}_{1/2}(^{76}\textnormal{Ge})=2.23^{+0.44}_{-0.31} \times 10^{25}$~yr, in two years of data with a 140~kg fiducial mass, EXO-200 would observe a signal of 91 (51) events above the projected background of 40 events, providing a significance of $8\sigma$ ($5.4\sigma$) using the most favorable (unfavorable) combination of $^{136}$Xe and $^{76}$Ge matrix elements for the translation between the two isotopes and the upper (lower) bound of the $^{76}$Ge interval.  Again, the smaller active mass implies that such sensitivity will be reached in a longer period of time.  EXO-200 is expected to run in low background conditions for up to four years.

EXO-200 has already made the first observation of $\tnbb$ of $^{136}$Xe, measuring a half-life $T_{1/2} = (2.11 \pm 0.04 \textnormal{ stat} \pm 0.21 \textnormal{ sys}) \times 10^{21}$ years~\cite{Ackerman}.  The nuclear matrix element corresponding to this half-life is 0.019 MeV$^{-1}$, the smallest measured to date. At $\znbb$ sensitivities discussed above, $\tnbb$ produces a negligible background.

\subsection{Xenon supply}

The choice of $^{136}$Xe for the EXO program derives from several considerations:
\begin{itemize}
\item The $^{136}$Xe Q-value of 2457.8 keV~\cite{Q-value} is large and above the energies of most \gr s from naturally occurring radionuclides, with the notable exception of the 2615~keV \gr\ from $^{208}$Tl.
 \item The double beta decay source and detection medium are the same material.  This configuration minimizes background and energy loss of the decay electrons.  For this reason, this is likely to be the only practical option for building very large detectors.  While the double beta decay event in the LXe appears as a localized energy deposition, most \gr\ interactions involve Compton scatterings that can be identified, in a large homogeneous detector, by their multi-site energy deposition. 
 \item The use of a material in the form of a liquid or gas allows for easy transfer of the enriched isotope from one detector to another.   In addition, the possibility of using the material either in gas or in liquid phase, with complementary properties, opens a broad set of possibilities for a program of experiments in which a large fraction of the cost is the isotopic enrichment.     
 \item $^{136}$Xe is particularly economical to enrich from the natural fraction of 8.9\% because it is a gas at standard temperature and pressure and hence easy to process in ultracentrifuges.    
 \item In a large detector, the xenon can be continuously re-purified, if necessary, during the lifetime of the experiment.     This is important because the background requirements for the next generation of $\znbb$ experiments are so extreme that the final detector is essentially the only device with sufficient sensitivity to verify the purity of the source.  Because it is a noble gas, Xe is particularly  easy to purify from all chemically active contaminants.    
 \item No long lived radioisotopes of Xe exist, hence, after a short ``cool down'' period underground and chemical purification, no contamination should remain in the gas or liquid.
 \item The double beta decay of $^{136}$Xe produces a barium ion ($^{136}$Ba$^{2+}$) that can in principle be tagged, drastically reducing backgrounds.  The techniques for Ba tagging are being developed by the EXO-collaboration but are not part of the design of EXO-200.
\end{itemize}

\begin{figure}
	\centering
	\includegraphics[width=4in]{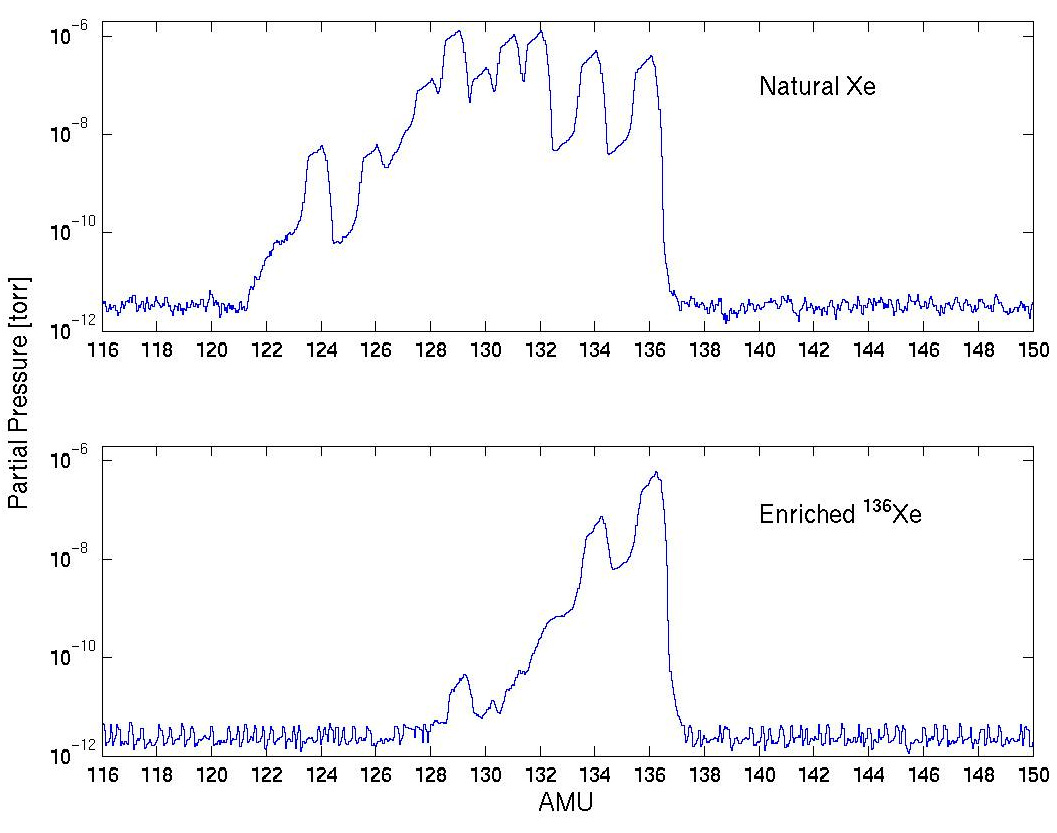}
	\caption{Mass spectra of natural Xe (top) and the EXO-200 enriched Xe (bottom), as determined by a residual gas analyzer.  $^{136}$Xe 	($^{134}$Xe) is present at 80.6\% (19.1\%) number concentration while the concentration of other isotopes is negligible.}
	\label{fig:enrXe}
\end{figure}

200~kg of xenon enriched to 80.6\% in the isotope $\xe$ (number concentration) were produced by ultracentrifugation starting from natural feed stock.  The mass spectrum of the material is shown in Figure~\ref{fig:enrXe}.  19.1\% of the atoms in the enriched xenon are of the isotope 134, and the rest of the natural isotopes are present in negligible concentration.   The isotopic enrichment was carried out in two installments by several laboratories in Russia using clean centrifuges, after which the gas was cryopumped into ten electropolished stainless steel cylinders.  This procedure was intended to guarantee chemical purity.  However, during the filling of EXO-200 it was found that at least one of the ten cylinders was contaminated with some heavy, fluorinated molecule.   This substance may be a lubricant used in ultracentrifuges~\cite{centrifuge}.

The detector was commissioned using a supply of natural xenon.   Measurements performed using an enhanced mass spectroscopic method~\cite{Leonard:2010zt} found that the natural xenon had a contamination of $42.6\pm5.7$~ppb~(g/g) of Kr~\cite{Dobi:2011zx}.   Although $^{85}$Kr is not a background for $\znbb$, it is a background for low energy processes, particularly problematic for dark matter detectors.  As expected, the Kr contamination in the enriched xenon is drastically lower at $25.5\pm3.0$~ppt~\cite{Dobi:2011zx}.

\section{The EXO-200 TPC}

The TPC is designed for the collection of both scintillation and ionization signals.  The LXe vessel volume of 58~liters is set by the total amount of enriched Xe available.  This volume includes the Xe and cable conduits and the high voltage (HV) feedthrough, shown in Fig~\ref{fig:TPC}, that are full of LXe during normal operations.  The total mass of LXe in EXO-200 is $\sim175$~kg, with 110~kg available in the active detector volume.  The remaining mass of Xe is in the gas handling system.

\begin{figure}
	\centering
	\includegraphics[width=5.0in]{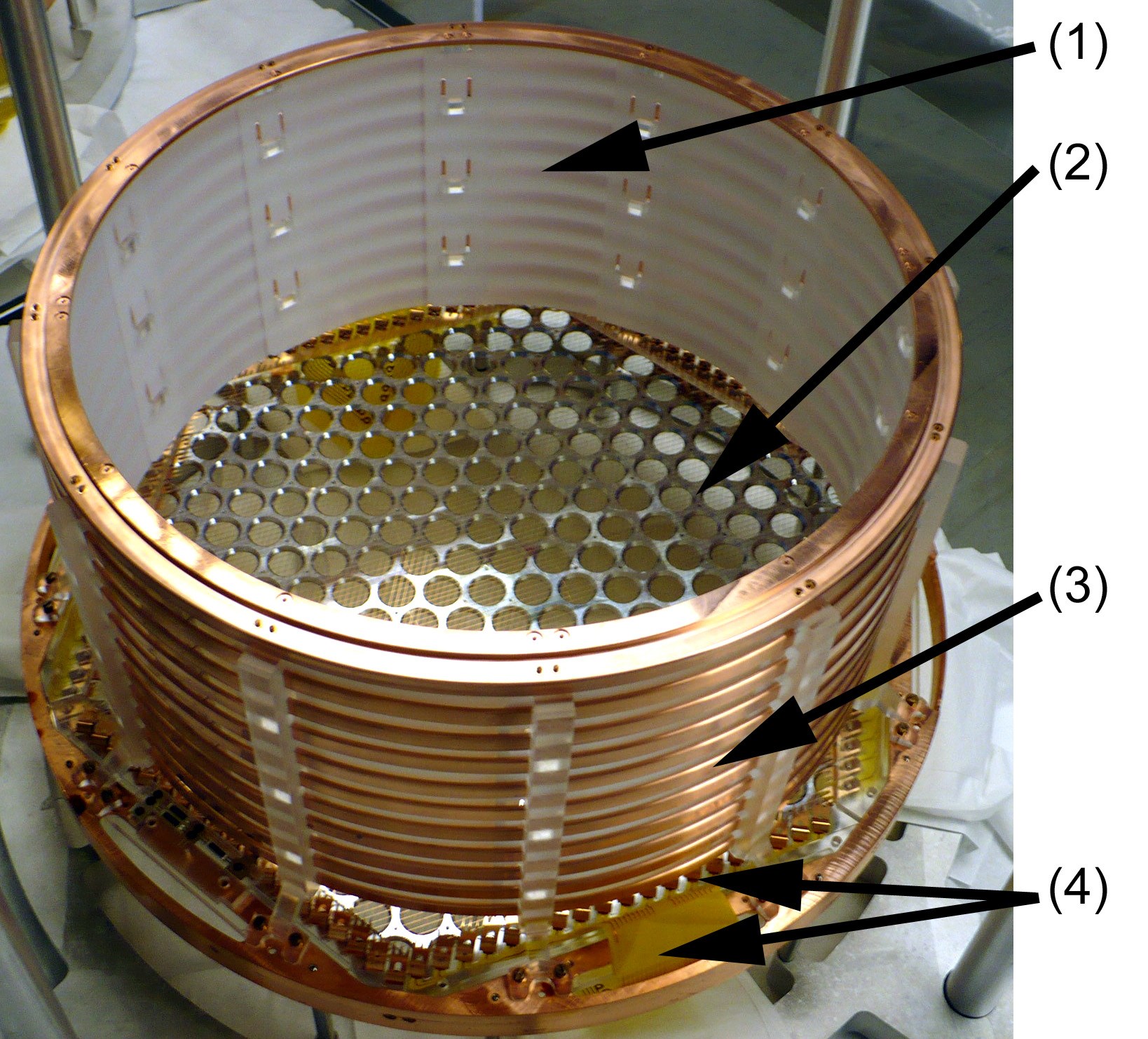}
	\caption{A view into the active Xe volume of one of the two EXO-200 TPC modules. PTFE tiles (1) installed inside the field-shaping rings serve as reflectors for the scintillation light. The aluminum-coated side of the LAAPD platter (2) is visible, as well as the field cage (3), ionization wires, and flexible cables (4).
	\label{fig:teflontiles}}
\end{figure}

Radiation depositing energy in the liquid xenon creates both a scintillation and an ionization signal.  The scintillation is detected almost instantaneously by the avalanche photodiodes, while the ionization is drifted in a uniform electric field to the crossed wire planes.  The TPC is divided into two almost identical drift regions, one of which is shown in Figure~\ref{fig:teflontiles}.    A cathode grid, held at negative high voltage, separates the two regions while the readout planes at both ends are held at ground potential.  This arrangement reduces by a factor of two the drift voltage at the expense of very little extra material (the cathode grid) in the middle of the active Xe volume.
 
The TPC axis is horizontal to provide identical signal path-lengths from both ends and to simplify the cathode feedthrough.   The cryostat axis is horizontal (but perpendicular to the axis of the TPC) because of limitations in the overhead space at WIPP.    This results in the TPC being cantilevered off the inner (cold) hatch of the cryostat.

\subsection {Scintillation channel}

The collection of the 178~nm scintillation light in EXO-200 is accomplished using 468 LAAPDs. Compared to photomultiplier tubes, LAAPDs have substantially lower radioactivity content, occupy less space, are intrinsically compatible with cryogenic operations, and have higher quantum efficiency at 178~nm.  The disadvantages of lower gain and higher noise are tolerable at LXe temperature and for the relatively high energies of interest here.  To fully exploit the LAAPD advantages, in EXO-200 they are used as bare dies, without standard ceramic encapsulation and directly mounted on two specially designed support platters by means of springs.   Light collection efficiency is improved through the use of reflective PTFE tiles around the field cage, where the electric field configuration makes it impractical to install LAAPDs.  Since the cathode grid has a high (90\%) optical transparency, LAAPDs mounted at each end of the TPC detect scintillation produced anywhere in the active LXe.

The EXO-200 LAAPDs are the unencapsulated version of Advanced Photonix part number SD630-70-75-500 and are described in detail in~\cite{ourapds}.  Each LAAPD has a 16~mm diameter active area (200~mm$^2$), with an overall diameter between 19.6~mm and 21.1~mm. The thickness of each varies between 1.32~mm and 1.35~mm.  The devices consist of a p-type epitaxial layer grown on n-type neutron transmutation doped silicon.  A second silicon wafer, ring shaped and gold coated, is bonded by a layer of aluminum to form the anode.  The external contact of the cathode is also gold coated. The edge of the device is then beveled and coated with a polyimide film to improve breakdown and dark current characteristics.  All LAAPDs were manufactured in a class 1000 cleanroom and stored in static dissipative boxes under nitrogen atmosphere from production to installation.

A total of 851 LAAPDs were purchased and delivered over the course of two years.  The gain, relative quantum efficiency, and noise characteristics of all purchased LAAPDs were measured at LXe temperature~\cite{Neilson:2009kf}.   At 1400~V bias, the LAAPD capacitance was measured to be 125~pF.  Of the devices that met the operating specifications, the 468 LAAPDs exhibiting the lowest noise were chosen for installation. 

\begin{figure}
	\centering
	\includegraphics[width=6in]{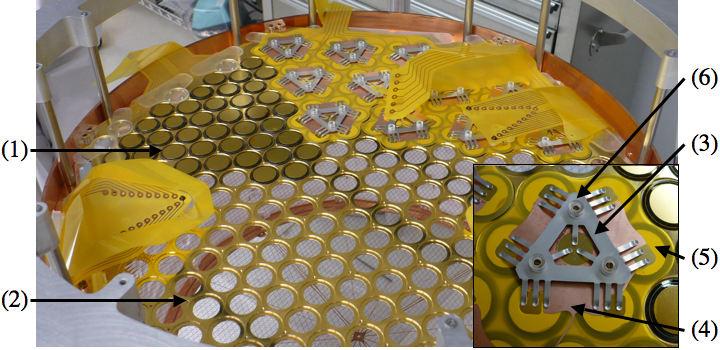}
	\caption{Bare LAAPDs are placed in a sector~(1) of one platter~(2).  The platter is gold coated on this side for improved electrical contact.   Platinum plated photo etched phosphor bronze ``spider'' springs (3) anchor gangs of 7~LAAPDs to the copper platter, and provide electrical contact between the cathodes of each LAAPD and copper traces~(4) on flexible cables~(5), as shown in the inset. Acrylic washers~(6) prevent electrical contact between the springs and the mounting screws that are threaded into the platter. Ionization-collecting wires (and the cathode grid) can be seen through the partially filled LAAPD platter.
	\label{fig:PlatterWGO7}}
\end{figure}

Two LAAPD platters, machined from the same low-activity copper stock as the LXe vessel, provide both mechanical support and common electrical potential for the 468 LAAPDs.  Their front sides are vacuum coated with aluminum and MgF$_2$ to reflect the scintillation light impinging between LAAPDs.  On the reverse side, the platters are gold coated to improve electrical contact with the front electrodes (anodes) of the LAAPDs.  Each platter (see Figure~\ref{fig:PlatterWGO7}) holds 234 LAAPDs in a canted hexagonal pattern with a photosensitive packing ratio of 48$\%$. 

The vacuum coating of the platters was performed by VPE Inc.~\cite{vpei}.  A 30~nm thick nickel layer was deposited under the 100~nm thick gold to prevent diffusion.  On the front side the 100~nm thick aluminum coating was deposited directly on the copper and then over-coated with a 50~nm thick layer of magnesium fluoride to prevent aluminum oxidation that would reduce the reflectivity at 178~nm.  The gold, nickel, and magnesium fluoride were purchased from Cerac Inc.~\cite{cerac} and, like all other materials, radioactivity certified.   The Al stock was the same as that used in LAAPD production and described in ~\cite{Neilson:2009kf}.

The anodes of all 234 LAAPDs are electrically connected together and held at a common voltage, -1400~V on one platter, and -1380~V on the other.  The platters were fabricated to accommodate 259 LAAPDs, though after fabrication it was decided not to install some of the  devices at the edge of each platter as their fields of view are either blocked or outside  of the field cage.   These empty locations were instead fitted with acrylic insulators designed to prevent electrical shorts.

The 12 ``tiles'' (see Figure~\ref{fig:teflontiles}) mounted inside the field cage to reflect the scintillation photons impinging on that region were made out of 1.5~mm thick skived TE-6472 modified PTFE from DuPont~\cite{dupontptfe}, specially sintered using a clean process by Applied Plastics Technologies~\cite{apt,LePort:2006bf}.  The inner radius of this PTFE lining is 18.3~cm.  Measured in Ar gas at room temperature, the reflectance for 178~nm light is 70$\%$ for the TE-6472 (expected to be higher in LXe), as opposed to $5-10\%$ for copper, depending on the oxidation level.

Each of the two LAAPD planes is divided into six sectors, four containing 38 LAAPDs each and the remaining two containing 37 and 45 devices.  In the special sector containing 37 LAAPDs one device is replaced by a PTFE diffuser 20~mm in diameter and 0.7~mm in thickness.   Three redundant bare, multimode optical fibers deliver light from an external laser pulser to the diffuser, which is used to test basic functionality and stability of the LAAPDs during detector operation.

In order to reduce the number of connections, LAAPDs are read out in ``gangs'' of five to seven devices, connected together by the ``spider'' springs shown in Figure~\ref{fig:PlatterWGO7}.  The springs, photoetched from 0.25~mm thick phosphor bronze and platinum plated, have multiply-redundant fingers contacting the cathodes (back sides) of the LAAPDs and doubling as mechanical restraints.   A special set of fingers for each spider connects to a copper trace on a flexible polyimide circuit bringing the signals outside of the TPC.    As discussed in~\cite{Neilson:2009kf}, the LAAPDs exhibit a substantial spread in the voltage required to achieve a fixed response (defined for this purpose as gain $\times$ relative quantum efficiency).  This is compensated for by loading each sector with LAAPDs of similar performance and providing a separate {\it trim} voltage to each of the sectors.   This trim voltage is applied to the cathodes through the same traces extracting the signals from the various gangs.   This system reduces the overall spread of response to $\sigma =2.5$\%.

\subsection{Ionization channel}

The ionization channel includes the TPC cathode and field cage, defining a region of uniform electric field, the charge collection (U) wire plane, sitting at virtual ground, and a shielding (V) wire plane located in front of the U wires and biased to ensure full electron transparency.   The V wires are also read out, and their inductive pickup signals provide a second coordinate without an additional shielding wire plane.   Both U and V wires are in front of the LAAPDs, and each has 95.8\% optical transparency.  The use of only two wire planes reduces the material budget of the TPC, increases the optical transparency, and maximizes the fiducial volume.

\subsubsection{Wire readout planes}

Both the cathode plane and the wire readout planes are made by photoetching sheets of phosphor bronze.  U and V wire planes are spaced 6~mm from each other and each is comprised of an array of parallel wires.    U and V wires in each plane are oriented 60$^{\circ}$ from each other, allowing two-dimensional reconstruction of the ionization cloud location.

\begin{figure}
	\centering
	\includegraphics[width=4in]{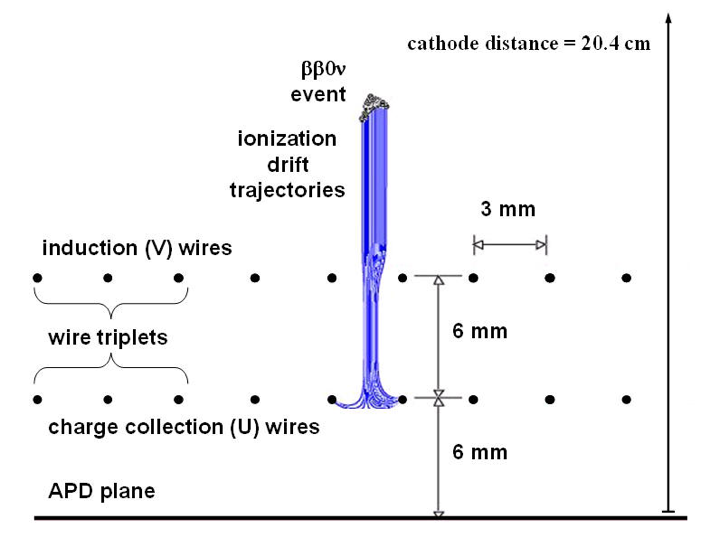}
	\caption{Geometry of the readout wire planes also showing a simulation of an electron cloud. The anode (U) and induction (V) wires, shown as collinear here, are in reality at 60$^{\circ}$ from each other.
	\label{fig:driftelectrons}}
\end{figure}

The distances between the two wire planes and the APD platter (6~mm each) and the wire pitch (3~mm) were chosen to maximize the fiducial volume while not requiring excessive high voltage on the V wire plane to achieve electrical transparency. The field uniformity, capacitance, and electrostatic forces for this configuration were evaluated with the MAXWELL~\cite{maxwell} electric field solver (see Figure~\ref{fig:driftelectrons}), and were found to be acceptable. Assuming a circular wire diameter of 125~$\mu$m, full electric transparency of the V wires is expected to be obtained when the U-V field is 131\% of the drift field~\cite{Bunemann}. MAXWELL confirms that the rectangular wire cross section employed by EXO-200 behaves similarly with respect to electrical transparency. In practice, however, an electric field ratio approaching 200\% was adopted to allow for mechanical tolerances and the deflection of the wires under the action of the electric fields.

Further optimization of detector performance leads to a {\em readout} pitch that is substantially larger than the 3~mm required by the field optimization.   This is because more electronics channels increase complexity and materials near the fiducial volume and do not necessarily provide better topological discrimination between signal (single-site events) and background (dominantly multi-site events from Compton scattering of \gr s).   The readout spacing is set to 9~mm, leading to triplets of wires with 3~mm pitch. This results in 38 read out channels for each of the two U and V wire planes at each end of the TPC, with a capacitance of 0.51~pF/cm between the anode triplets and ground.

\begin{figure}
	\centering
	\includegraphics[width=6in]{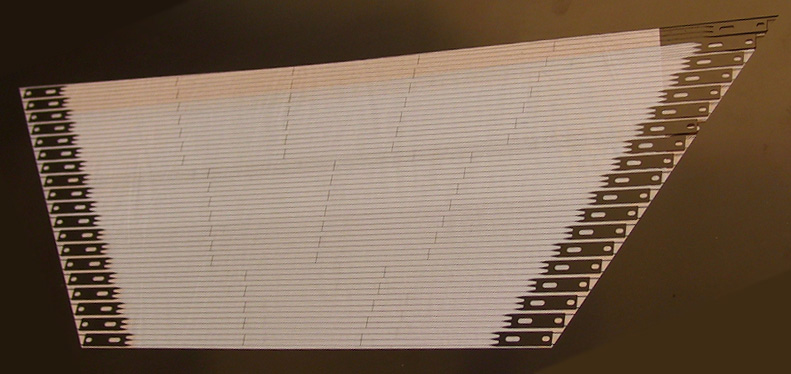}
	\caption{A panel of photoetched phosphor bronze wires.  The panel contains wire triplets for half of one U or V wire plane.
	\label{fig:wirepanel}}
\end{figure}

\begin{figure}
	\centering
	\includegraphics[width=6in]{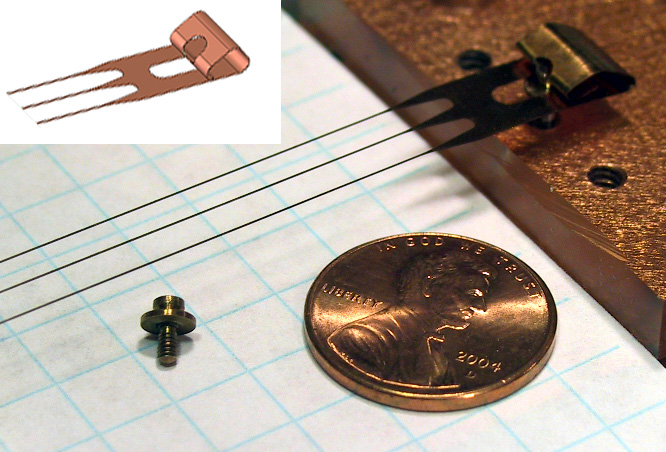}
	\caption{A wire triplet installed on its support screw after forming the spring (in the actual detector, the screws are threaded on acrylic supports).  The screw is  custom designed size 0-80 UNF, made out of phosphor bronze.   The inset shows the spring folding scheme.
	\label{fig:wiredesign}}
\end{figure}

The wire triplet arrangement was found to conveniently solve both the problem of mounting the wires in a compliant manner, suitable for high reliability under thermal cycling, and the desire for low radioactivity content.   Triplets were photoetched with a clean process by Vaga Industries~\cite{vaga} from panels of 0.13~mm thick, ``full hard'' CA-510 Grade A phosphor bronze, as shown in Figure~\ref{fig:wirepanel}.  The stock material was obtained from E.~Jordan Brookes~Co.~\cite{ejbco} and qualified for radioactivity.    A suitable spring is then obtained by folding the ends of each triplet, as shown in Figure~\ref{fig:wiredesign}.    The springs behave linearly for tensions less than 8.8~N with spring constant 28.9$\pm$0.7~N/cm, and begin to yield at 9.8~N.  While tensions in the allowed range for this spring are sufficient to ensure wire stability at the electric fields of interest, ``bridges'' connecting the wires in triplets are provided every $\sim$10~cm, as shown in Figure~\ref{fig:wirepanel}.   Individual wires in each triplet have a roughly square cross section with width of 127$\pm$40~$\mu$m due to tolerances in the photoetching process.    The unusual shape of the wires can be tolerated because the detector is operated at unity gain. 

\begin{figure}
	\centering
	\includegraphics[width=6.0in]{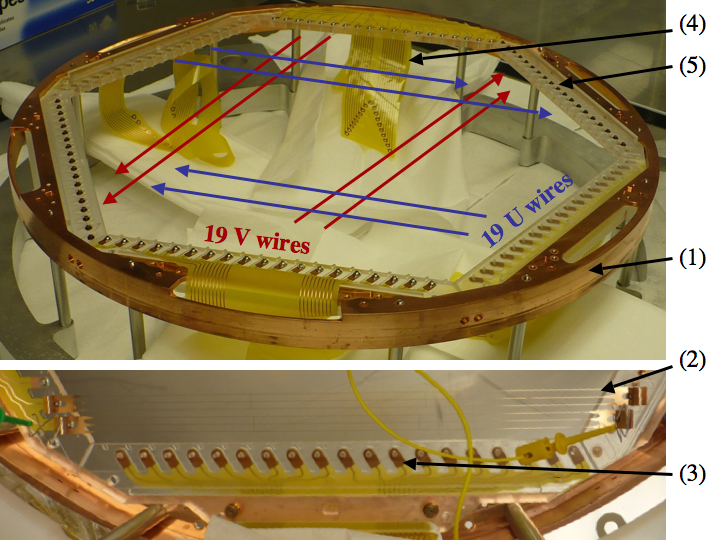}
	\caption{A copper support ring (1) holds six acrylic blocks in a hexagonal pattern. U wires (2) are mounted on one side of the acrylic blocks and V wires (not shown) are mounted on the opposite side (3) providing a spacing of 6~mm between the wire planes. Four flexible cables (4) make the electrical connections to platinum plated 0-80 UNF screws which anchor the wire triplets to each of four of the acrylic blocks. Un-plated 0-80 UNF screws (5) serve to anchor the other end of the wires and are not used for electrical connection.
	\label{fig:wiresupport}}
\end{figure}

Custom 0-80 UNF size screws on either end of a triplet anchor the spring connections to six 6~mm thick acrylic beams mounted in a hexagonal pattern onto a copper support ring, as shown in Figure~\ref{fig:wiresupport}.   Each U or V wire plane is stretched between two pairs of such acrylic beams, with the two wire planes mounted on opposite sides of the beams.   The thickness of the acrylic beams defines the 6~mm spacing between U and V planes.

Screws on one end of each wire triplet are platinum-plated to improve electrical contact and are tightened through the polyimide-based interconnect circuitry.  Platinum instead of gold plating was chosen in order to avoid potential $^{40}$K contamination that was found in the gold-plating solution.  The screws on the other end of each wire triplet are un-plated, minimizing the amount of plating in the detector.  As the wire triplets contact the screws at a sharp edge, the resulting contact force was deemed sufficient for the un-plated triplets.   The length of the 19 wire-triplets varies linearly from 22.8~cm to 41.5~cm (anchoring screw to anchoring screw) along each acrylic block. Due to the large coefficient of thermal expansion (CTE) of acrylic relative to copper and phosphor bronze (see Table~\ref{tab:ctes}) and the mounting geometry of the acrylic beams on the copper support ring, the shortest 28 wire-triplets in each plane experience an increase in tension during cooling, while the remaining 10 experience a decrease. The shortest triplets stretch by 0.28\%, and the longest shrink by 0.05\%, resulting in a change in tension during cooling ranging from approximately $+1.8$~N to $-0.6$~N.

\begin{table}[hbp]
\centering
\begin{tabular}{l c}
\lighttoprule \lighttoprule
Material & CTE [$\mu$m/(m$\cdot$K)]\\
\midrule
Copper & 16.9\\
CA510 phosphor bronze & 17.8\\
Acrylic & 73.6\\
\bottomrule
\end{tabular}
\caption{Coefficients of thermal expansion of materials relevant to ionization wire construction}
\label{tab:ctes}
\end{table}

Wire triplets were installed at room temperature with a tension between 2.9~N and 5.9~N, verified by measuring their resonant frequency when excited by an AC current in the static magnetic field provided by a large coil.    The resulting tensions at LXe temperature are sufficient to keep wire deflections below 0.1~mm for the maximum design voltage of 4000~V on the V plane.

\subsubsection{Field cage}

EXO-200 is designed to operate at a maximum drift field of 3.7~kV/cm.   The drift region is a cylinder of 18.3~cm radius and 38.4~cm length bound radially by the PTFE reflecting tiles installed inside the field shaping rings and longitudinally by the opposite V wires. 

\begin{figure}
	\centering
	\includegraphics[width=6in]{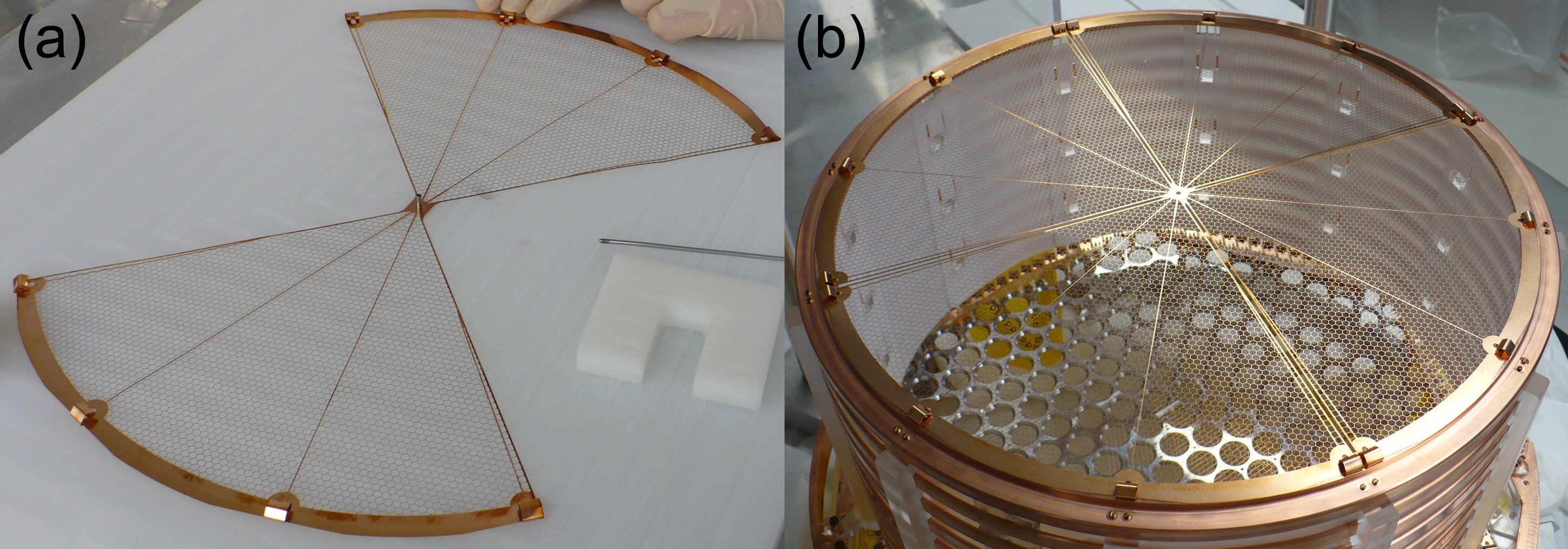}
	\caption{(a) One of the two phosphor bronze photoetched parts comprising the cathode. (b) The complete cathode installed in the detector.
	\label{fig:cathode}}
\end{figure}

The cathode is made out of the same phosphor bronze used for the wires, photoetched into a grid with 90\% optical transparency.   Because of the maximum width of the phosphor bronze stock, two half-cathodes are employed, as shown in Figure~\ref{fig:cathode}. The parts were photoetched at Vaga Industries in the same manner as the ionization wires, and utilize the same spring mechanism to maintain tension during thermal cycling. The cathode is mounted on the last copper ring of one of the two field cages.  Twelve custom screws, identical to those used to install the U and V wires, anchor the cathode to the copper ring. Four of these screws placed at opposite corners of each half-cathode are platinum plated to improve electrical contact. A mated copper ring, structurally completing the other field cage and electrically connected via platinum plated phosphor bronze leaf springs, is not loaded with grids, producing the only nominal asymmetry in the setup.

In each field cage the electric field is graded in ten steps by copper field shaping rings. Each ring is 0.97~cm long, and 37.4~cm in outer diameter, resulting in a radial clearance of 11 mm between the rings (at high voltage) and the copper vessel (at ground). The pitch between rings is 1.69~cm. Two dimensional~\cite{maxwell} simulations of this geometry, assuming exact cylindrical symmetry, predict full electron collection from a cylinder with radius 0.8~cm smaller than that of the field shaping rings.  This corresponds to a total active LXe mass of 110~kg.  

\begin{figure}
	\centering
	\includegraphics[width=6.0in]{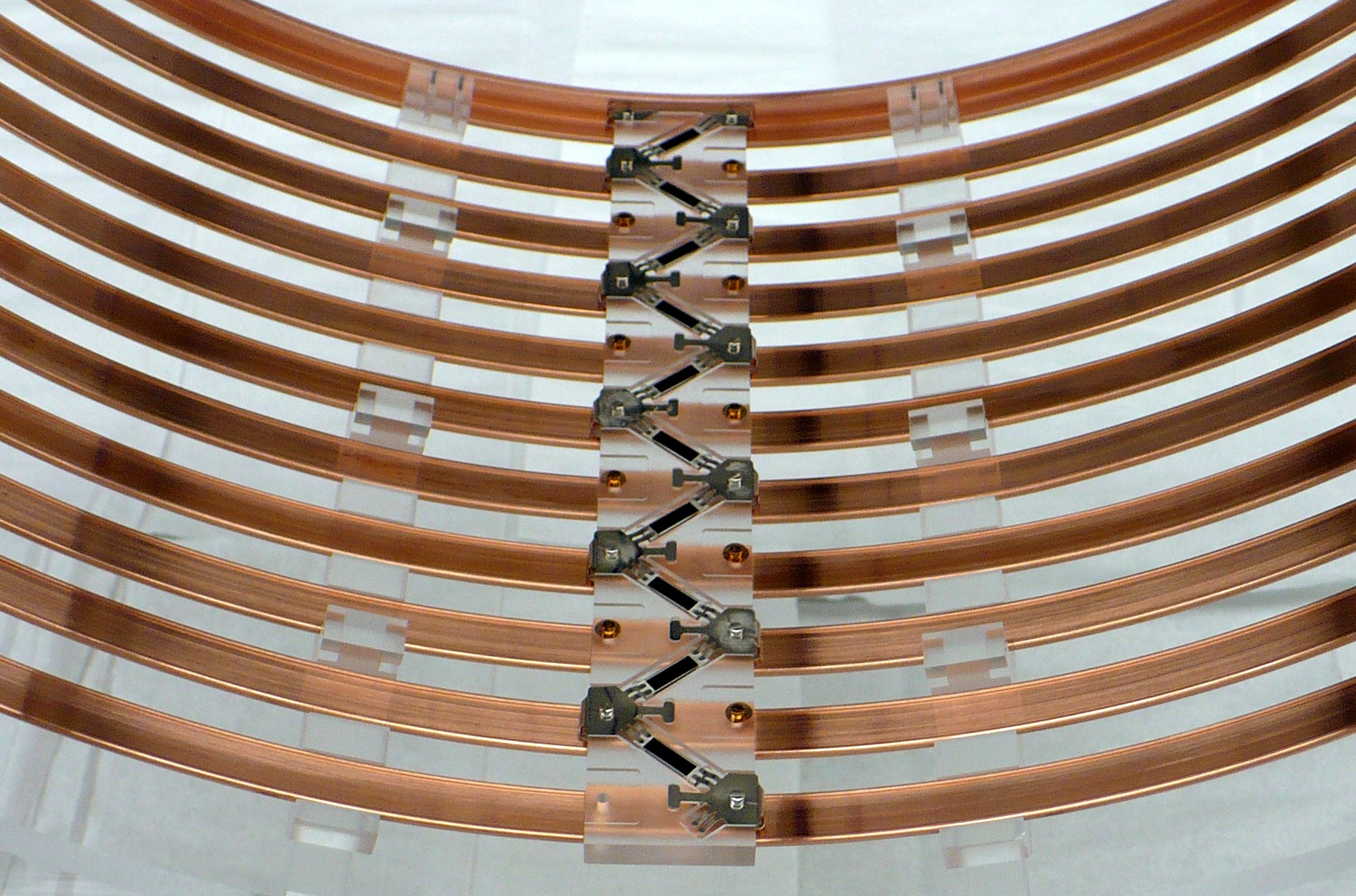}
	\caption{The resistor chain which grades the electric field in each TPC. Platinum plated phosphor bronze springs make contact between the resistors and field shaping rings. The chain and field shaping rings are mounted on acrylic ``combs'' (that also support the PTFE reflective tiles, not shown here).
	\label{fig:resistorchain}}
\end{figure}

The schematic drawing of the field grading circuit is shown in Figure~\ref{fig:electronics}.  Ten 900~M$\Omega$ resistors grade the potential between rings (see Figure~\ref{fig:resistorchain}), dissipating a total of 0.6~W at 3.7~kV/cm.  The V wires are connected to the last ring through 900~M$\Omega$ and to ground with 450~M$\Omega$, obtained as the parallel of two of the resistors above.   Individual V wires, that need to be read out, are further decoupled from each other by smaller-value resistors that are external to the chamber.    The 900~M$\Omega$ resistors are custom made using thick film technology on $3.3\times 13.1\times 0.9$~mm$^3$ sapphire substrates supplied by Swiss Jewel Co~\cite{swissjewelco}.   Resistors were fabricated by Piconics~\cite{piconics} using DuPont~\cite{dupontres} resistive and conductive pastes.    200 resistors were produced, from which two of the thirteen most closely matched sets were chosen for installation. The RMS dispersion for each of the two sets of resistors is 0.6\%.

The field cages including the resistors are supported by acrylic ``combs'' cantilevered off the copper support ring to which the wire and LAAPD planes are anchored. Contact with each field shaping ring and between resistors is made by platinum plated phosphor bronze springs, photo etched in the same manner as the ionization wires.  The PTFE reflector tiles are mounted on the same acrylic combs and cover the resistor chain.

\subsubsection{High voltage feedthrough}

Because of the background requirements, a special high voltage (HV) feedthrough to the cathode was developed for EXO-200.  

\begin{figure}
	\centering
	\includegraphics[width=6in]{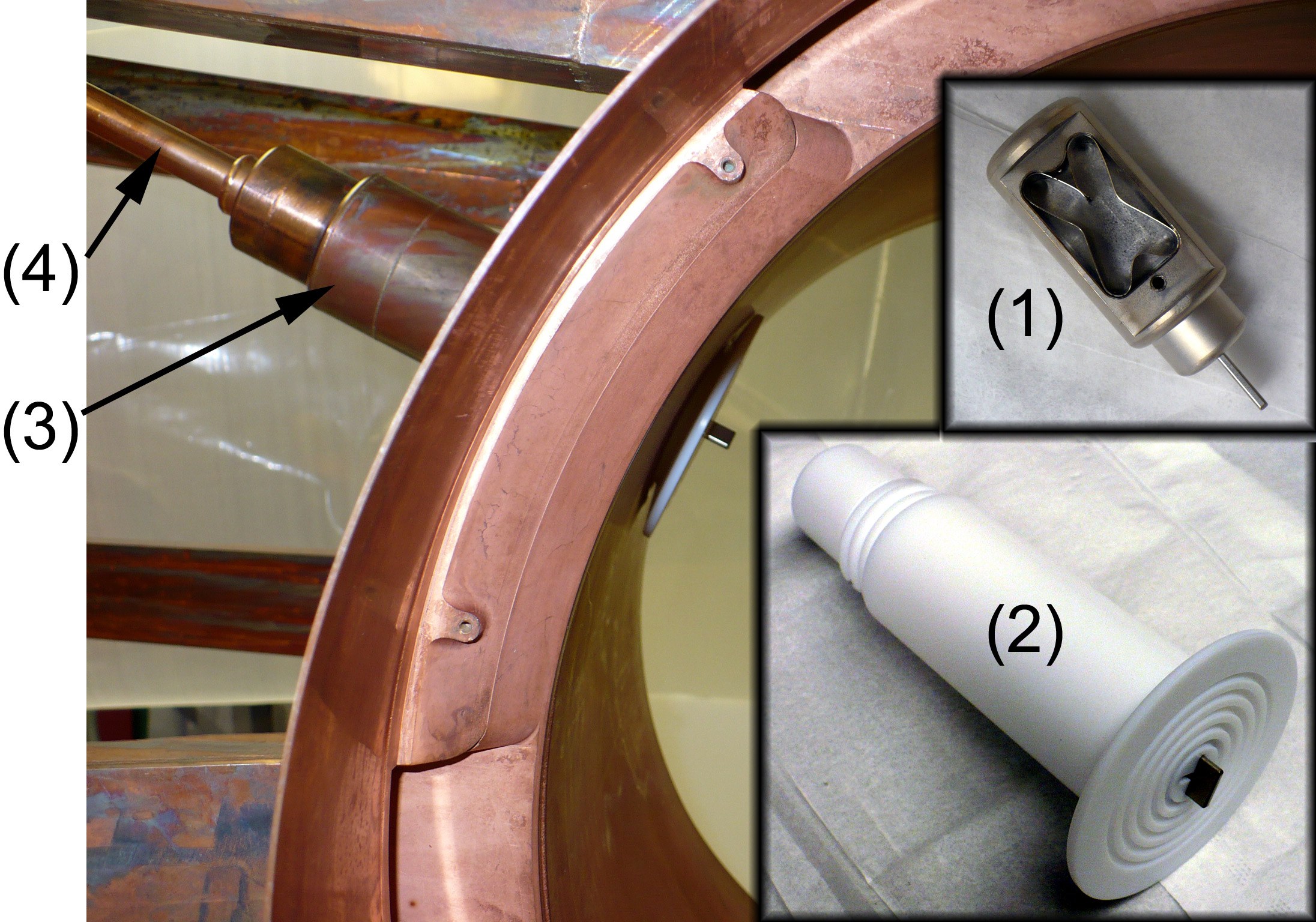}
	\caption{The high voltage delivery system: the phosphor bronze leaf spring in the copper receptacle (1).  Both components are platinum plated.   The receptacle is embedded in a PTFE block (2) and mounted in a special copper adapter (3) connecting the TPC body to the copper tube (4) guiding the high voltage cable from the outside.
	\label{fig:spud}}
\end{figure}

\begin{figure}
	\centering
	\includegraphics[width=6in]{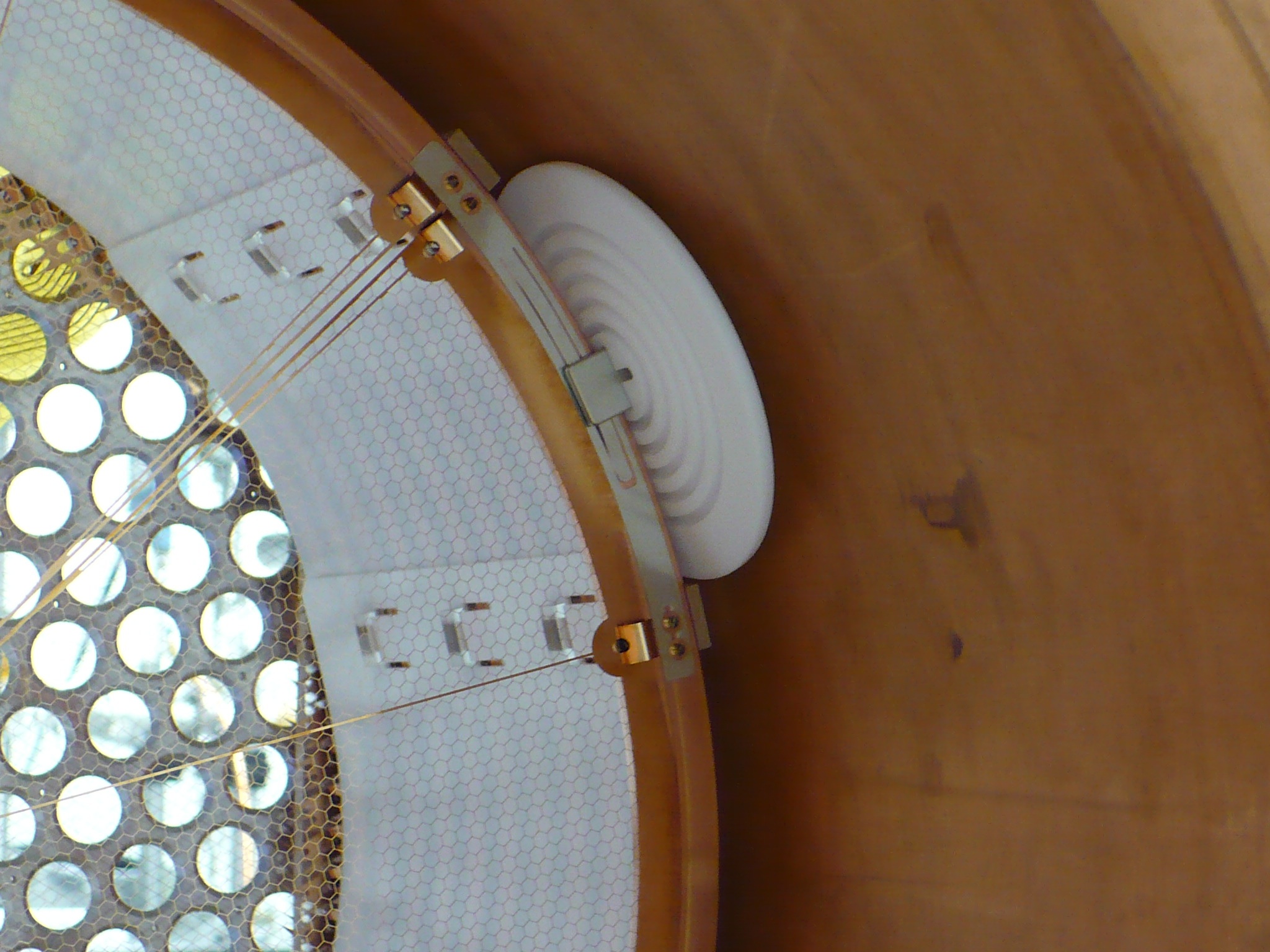}
	\caption{Platinum plated leaf spring connecting the HV feed to the cathode ring. The other field cage (with no cathode grid installed, is connected with a similar leaf spring.
	\label{fig:hvspring}}
\end{figure}

A HV delivery conduit is provided by a 5/8~inch copper pipe welded to the body of the TPC on one end, to the inner (cold) hatch of the cryostat and, after a jog, to the outer (warm) hatch of the cryostat.   This pipe effectively extends the TPC volume out of the cryostat, to room temperature, about 1.5~m away from the TPC.    The inner part (2.7~mm diameter solid copper conductor and 9.5~mm diameter polyethylene dielectric) of a RG217 coaxial cable (from Pasternack Enterprises~\cite{pasternack}) is inserted in this conduit, so that a final section of the cable, with the dielectric removed, makes contact with a spring-loaded receptacle embedded in a large PTFE block, as shown in Figure~\ref{fig:spud}.   The platinum plated receptacle and leaf spring are made, respectively, of copper and phosphor bronze~\cite{ejbco}, while the PTFE is the same DuPont TE-6472 material used for the reflective tiles.   A tab on the receptacle penetrates the PTFE block making contact with a platinum plated photo etched phosphor bronze leaf spring mounted on the cathode ring, as illustrated in Figure~\ref{fig:hvspring}.  

The HV delivery conduit is sealed at room temperature with a Swagelok compression fitting around the polyethylene dielectric of the cable.    Great care in making the Swagelok connection and the selection of a favorable region of cable are essential for obtaining a He-leak tight seal.   This arrangement also relies on the fact that, apparently, the polyethylene extrusion over the solid copper conductor provides a He-leak tight seal.    Xenon fills the space between the HV delivery conduit and the cable, in liquid phase at the low temperature and in gas phase at room temperature, with the liquid-gas interface at some intermediate location.  The polyethylene of the cable dielectric was not baked and purged in dry nitrogen before use.    In order to flush away oxygen desorbing from the plastic, a connection was provided at the room temperature end of the feedthrough so that efficient pumping and recirculation could be established.   

The polyethylene and copper of the HV cable were certified for radioactivity content like all other components.   The total mass density of the cable is 1.11~g/cm, with 59.5~cm of it located within the volume of the cryostat inner vessel.

\subsubsection{High voltage instabilities}
While the entire HV system was nominally designed to work up to 75~kV (3.7~kV/cm field), operations above 9 to 14~kV have shown instabilities that manifest themselves as $\sim$mV-scale glitches on the HV feed line.   Various tests appear to hint that the problem is due to some corona discharge, possibly on the sharp edges of various photoetched parts.  Although conditioning cycles may heal this problem, the physics limitations of a lower field are not regarded as sufficient to justify the risk associated with conditioning.  The HV system is operated stably at 8 kV, and it is unlikely that a better understanding of the glitch phenomenon will be available until a substantial low background data set is in hand.

\subsection{Detector wiring}

Electrical connections are a substantial challenge in the EXO-200 design.   For reasons of practicality the decision was made to use warm electronics behind the HFE-7000 and the lead shielding, about 1~m away from the readout wires and LAAPDs.   This choice eliminated the effort of designing low background electronics at the expense of substantial wiring. Traditional connections, including solder joints, were deemed too radioactive and a potential risk of xenon contamination with electronegative impurities. Instead, electrical connections were made using 18~$\mu$m thick copper traces on flat, 25~$\mu$m polyimide flexible cables.   The cables were photo etched from adhesive-free polyimide flexible copper clad laminate~\cite{holders} purchased through Nippon Steel Chemical Co.~\cite{nippon}.   The same technique was used for two types of interconnections: cable panels were used to group and route signals and bias voltages in the pancake-shaped volume behind the LAAPDs and long cable strips are used to bring signals out and bias voltages to the TPC via six rectangular copper tubes (referred to as ``legs'') welded between the inner hatch of the cryostat and either end of the TPC vessel.   The legs also support the TPC from the inner cryostat hatch and, in two cases, are used to pump down the TPC and feed and recirculate the xenon.  On each end of the TPC, flat cables servicing the U wires, V wires and LAAPDs travel in separate tubes.  Connections to the copper traces are made using platinum plated phosphor bronze springs or 0-80 UNF screws pressing together copper traces.    

\begin{figure}
	\centering
	\includegraphics[width=6in]{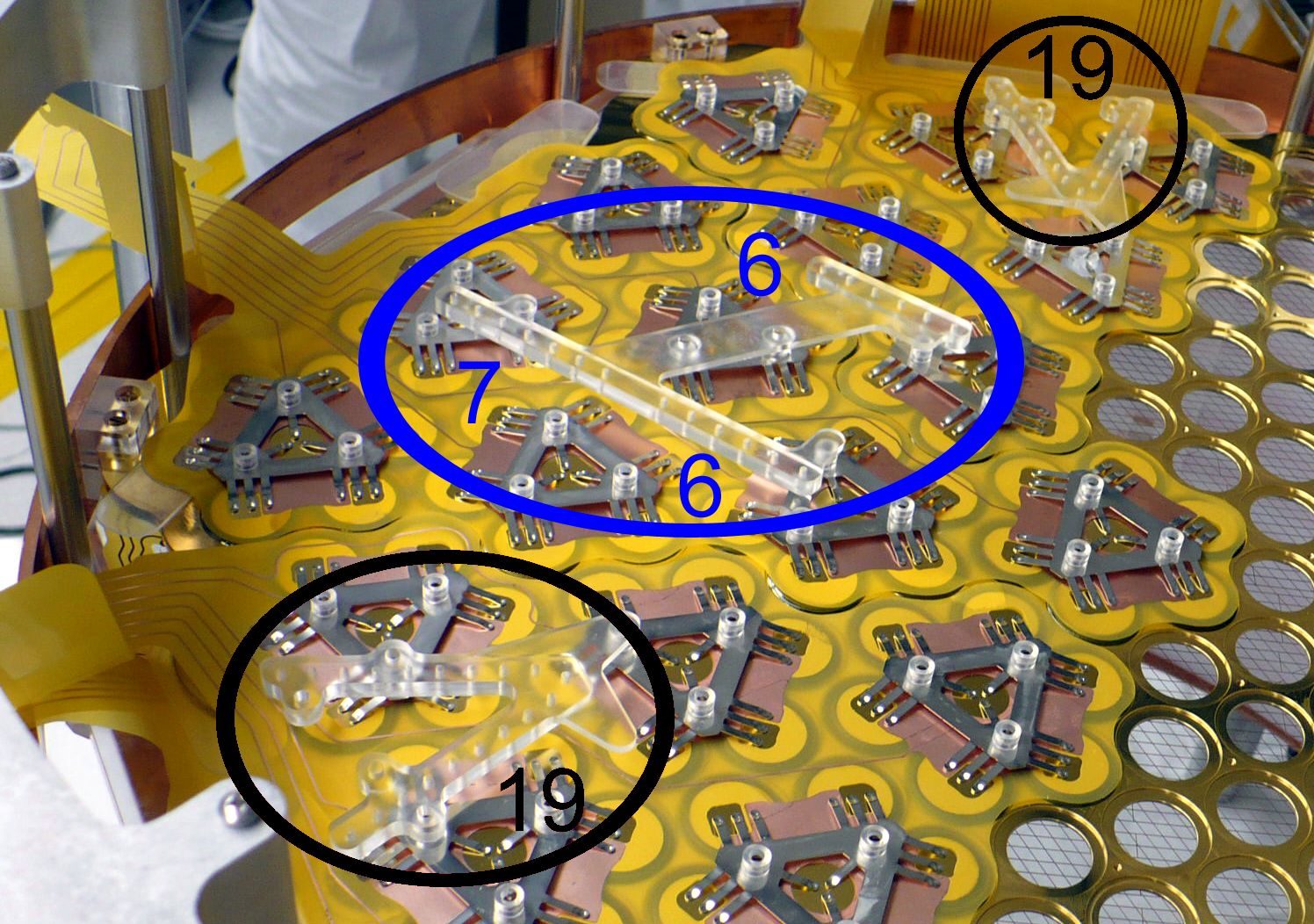}
	\caption{Acrylic ``Tee'' blocks (indicated in blue) and ``Vee'' blocks (indicated in black) used to mate panels with long cables. The number of channel connections made on each block is indicated.
	\label{fig:teesvees}}
\end{figure}

Each of the four U and V wire planes makes contact with two short flexible cables, each carrying 19 channels (see Figure~\ref{fig:wiresupport}). Each LAAPD plane makes contact to six short flexible cables, five carrying six channels and one carrying seven channels (see Figure~\ref{fig:PlatterWGO7}). The four pairs of ionization interconnects each mate with a cable on two acrylic ``Vee''-shaped blocks bolted to the LAAPD platter (see Figure~\ref{fig:teesvees}). Each of the two sets of six LAAPD interconnects similarly mate with an LAAPD cable on two ``Tee''-shaped acrylic blocks.  The mounting of the acrylic blocks is designed to permit thermal expansion.

\begin{figure}
	\centering
	\includegraphics[width=6in]{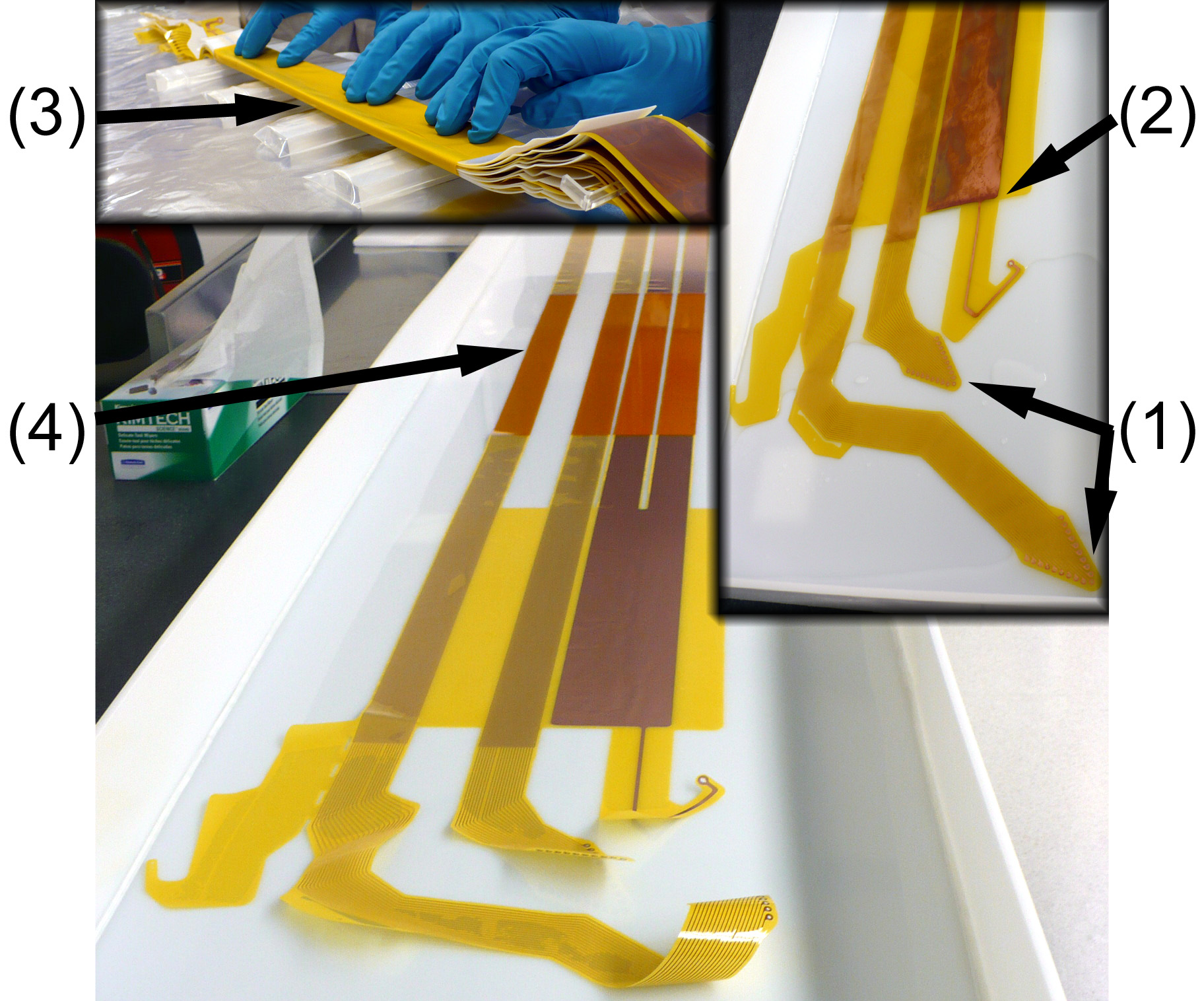}
	\caption{Examples of long cables. (1) Features at the head of the cable that mate with acrylic ``Vee'' blocks. (2) Conducting planes used to mitigate microphonic noise and improve electrical connections. (3) The cable bundled with PTFE insulators. (4) Coverlay applied to the portions of the cables outside of the inner vessel of the cryostat.
	\label{fig:xlongcable}}
\end{figure}

Some of the steps in the fabrication of one of the long cables are shown in Figure~\ref{fig:xlongcable}.   Each cable is made up of several strips that are then ``rolled'' into a cable bundle, appropriately shaped for insertion in one of the rectangular legs.   Since the 25~$\mu$m polyimide substrate cannot withstand the maximum voltage that may be applied between different layers (4~kV in the case of the V wires bias), sheets of 0.71~mm thick TE-6472 PTFE insulators are sandwiched in the cable folds.   

Some of the cable layers include traces while others are solid conducting planes serving a variety of purposes. In the case of connections to the U (charge collection) wires that sit at virtual ground, such planes provide 50~m$\Omega$ ground connections for the TPC.  In the case of connections for the V (induction) wires the conducting planes are biased at the same high voltage as the wires they are servicing, greatly reducing microphonics that would result from a vibrating conductor at high potential in close proximity to a ground.   Finally, in the case of the LAAPD cables, one conducting plane provides the $\sim$-1.4~kV bias to the support platter. Its large width provides a low inductance AC return path, required by the very short $\sim$50~ns rise time of the LAAPD signals. While such a plane has to run close to the LAAPD signal lines to reduce the effect of ground loops, a second plane at the LAAPD trim potential ($\leq$200~V) is located between the HV plane and the signal lines, again to reduce potential microphonic effects.

The trace patterns on the U and V wire cables are very similar.    Guard traces are interspersed between signal traces and differentially used in the front-end electronics.   The trace pitch is 1~mm, with a trace width of 0.5~mm.   There are 2~mm of bare polyimide beyond the edges of the outermost traces, for a total cable width of 43.5~mm.    The 37 traces on each LAAPD cable are split among three strips, and guard traces are omitted. This allows a trace pitch of 3~mm.

Trace resistance for all of the cables was measured to be $\sim$20~m$\Omega$/cm, resulting in a total resistance of $\sim$5~$\Omega$ for each trace. The capacitance between signal and guard traces on U and V cables was measured to be 30~pF.  Pyralux FR Coverlay~\cite{coverlay} was applied to the outermost  68.6~cm of each cable, starting after the cables exit the inner cryostat hatch feedthrough. These coverlays further insulate the conductor traces and increase cable robustness.

\begin{figure}
	\centering
	\includegraphics[width=6.0in]{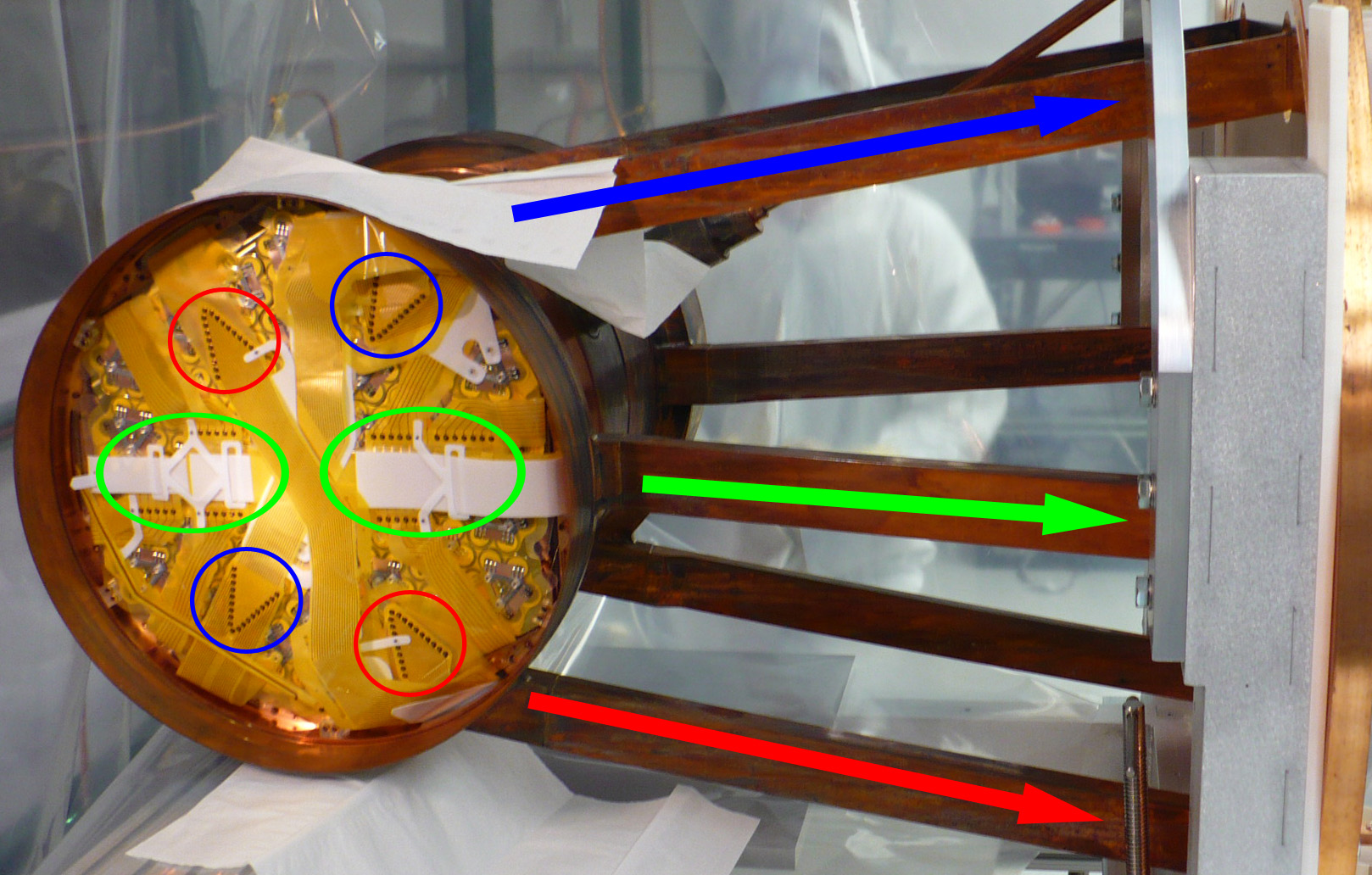}
	\caption{A view of one end of the TPC before sealing the copper vessel. The path and connections of V wire cables are indicated in red. Blue (green) indicates the path and connections of the U wires (LAAPD) cables.
	\label{fig:allcables}}
\end{figure}

\begin{figure}
	\centering
	\includegraphics[width=4in]{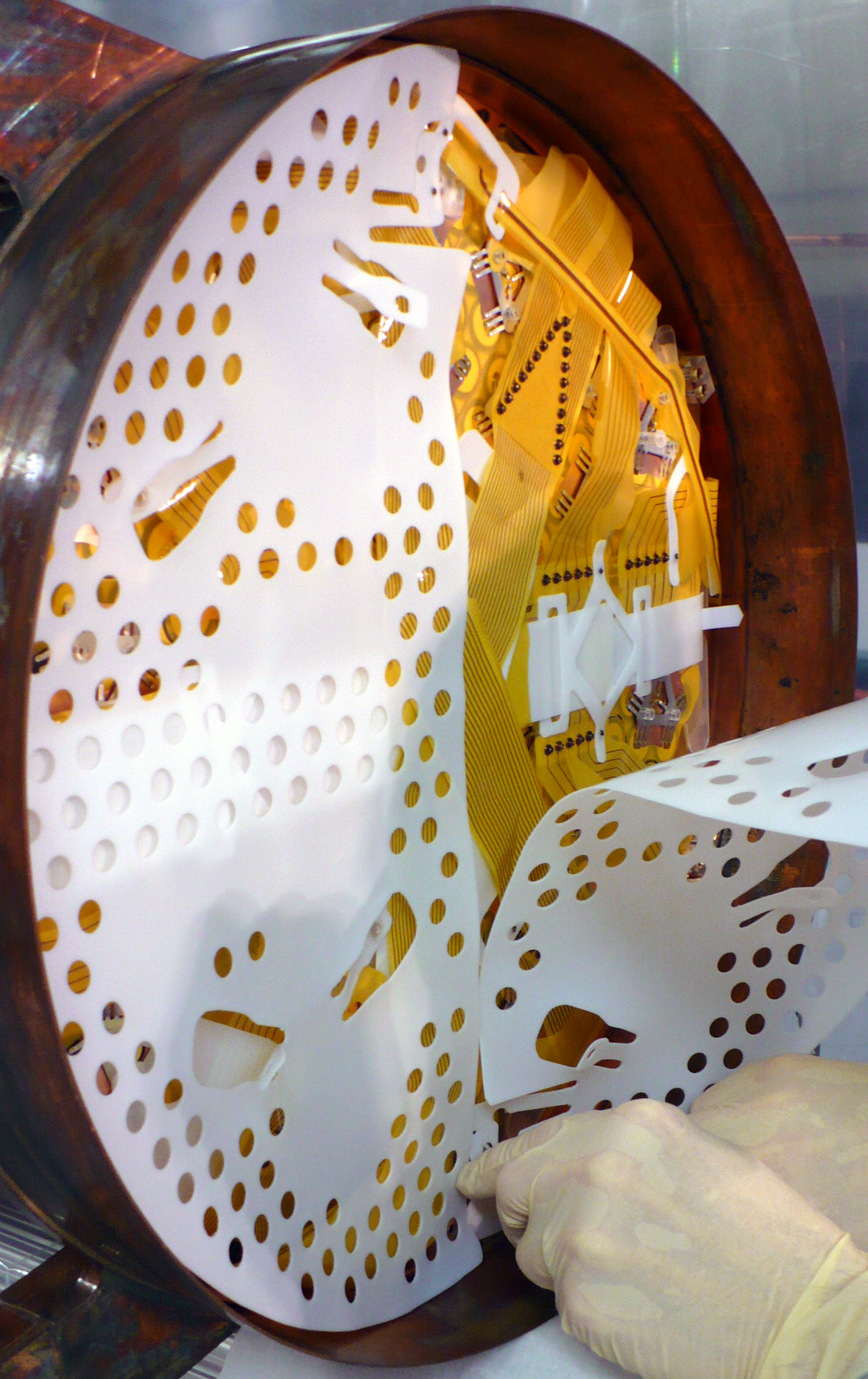}
	\caption{Two perforated PTFE half-circles separate the wiring system from the copper bulkheads.
	\label{fig:teflonprotectors}}
\end{figure}

Figure~\ref{fig:allcables} shows the back of one assembled detector end. The connections that the three cable bundles make at the back of the detector plane are highlighted. Also visible in the figure are additional PTFE protectors constraining the cables and interconnects and preventing electrical shorts.    A set of two PTFE half-circles is finally installed, as shown in Figure~\ref{fig:teflonprotectors}, further separating the wiring system from the copper bulkhead that is subsequently welded in place.  Holes in the PTFE half-circles improve LXe circulation and hence aid the purification process. To further reduce the risk of electric breakdown the copper legs containing the cables bundles are lined with 0.71~mm thick PTFE sheet.

All EXO-200 flexible cables were photoetched by FlexCTech~\cite{flexctech} under close supervision of EXO personnel stationed at the facility during the entire production.   Gloves, fresh chemicals, and new chemical containers were used, though cupric chloride, the principal etchant, was not replaced in the machines because of cost considerations. Several isopropanol rinses were added to the standard process and, in addition, all cables were subject to a post-production plasma etch.  Unlike all other components providing electrical contact, the flexible cables were not platinum plated, in consideration of the very delicate nature of the copper traces.   Cables were always stored in nitrogen boil-off atmosphere to prevent oxidation.

\begin{figure}
	\includegraphics[width=6in]{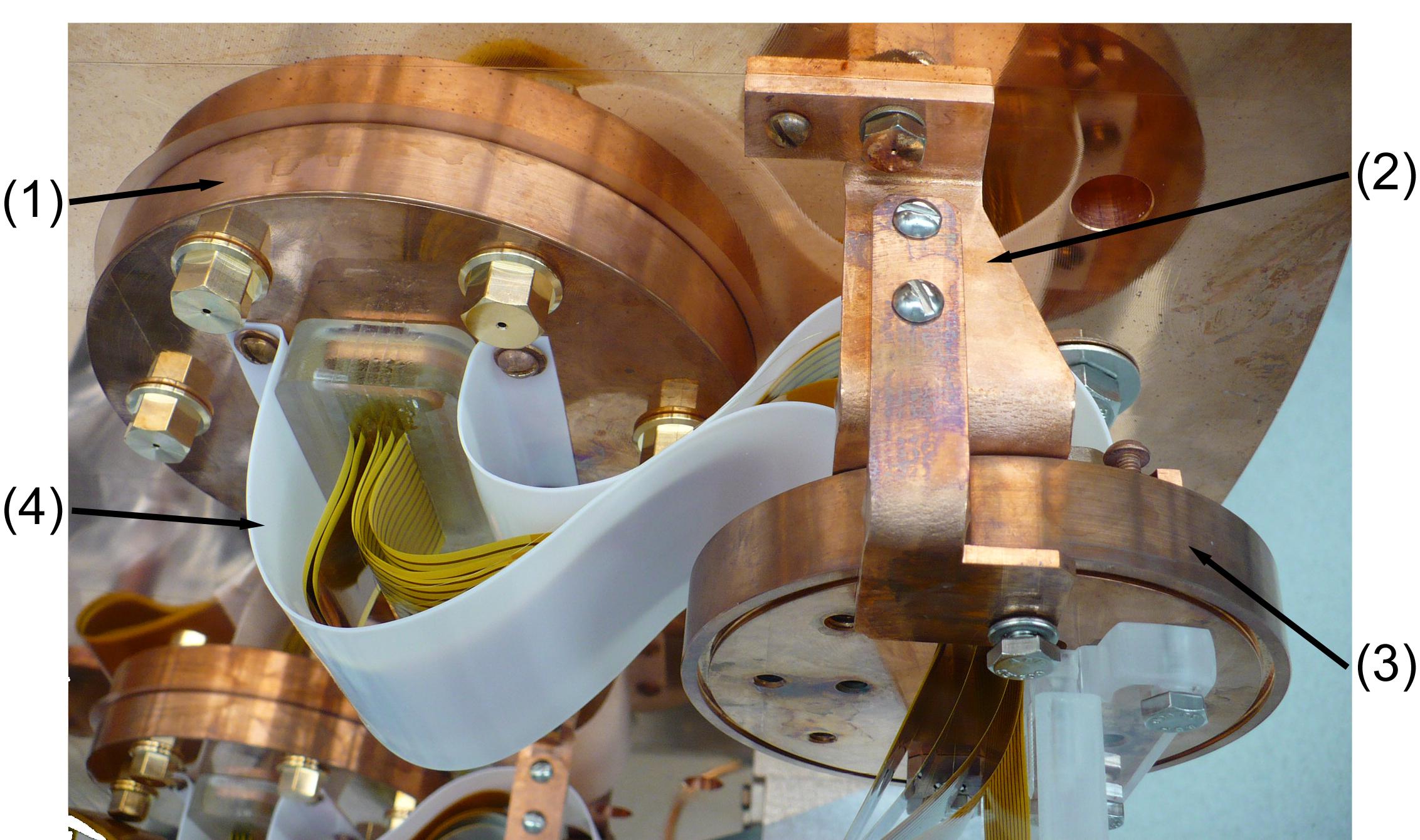}
	\caption{View of warm and cold cable feedthroughs, as installed. (1) Cold flange that makes the seal from LXe to vacuum, showing acrylic cup filled with epoxy. (2) Temporary bracket holding the warm flange during TPC transportation and installation.  (3) Warm flange making the seal from the cryostat insulation vacuum to atmosphere.  (4) PTFE strain relief.  The service loop shown is in the insulation vacuum of the cryostat.  The warm flanges are sealed against the inside face of the warm cryostat hatch.
	\label{fig:cablefeedthru}}
\end{figure}

Feedthroughs for the flexible cables transitioning from the LXe volume to the cryostat insulation vacuum and from the vacuum to the external atmosphere are formed in special copper flanges at the end of each leg (see Figure~\ref{fig:cablefeedthru}).   Two of the six flanges include a copper bypass for the circulation of the xenon.   On the cold (warm) hatch, the flanges are sealed using indium plated phosphor bronze sprung gaskets~\cite{jetseal} (silicone o-rings).  The cable feedthroughs are made using a low outgassing two component epoxy by Master Bond~\cite{masterbond}.   The fabrication of each feedthrough is a two step process. First, U-shaped acrylic parts are bonded to the cables using a very small amount of epoxy, forming a sealed cup. Liquid epoxy was then poured into the cup, bonding the cables and cup to a 0.5~mm thick copper lip on each flange.    Since the use of epoxies with thermal expansion properties matching those of copper is excluded by the high radioactivity content of the fillers surveyed, a substantial differential contraction at the feedthroughs is accepted by providing the thin and compliant copper lip.   

 \subsection{The TPC vessel}
A thin-walled quasi-cylindrical copper vessel houses the TPC and the LXe.  One of the design goals of this vessel was to maximize the usage of the enriched Xe.  Hence the central portion of the vessel, closely surrounding the field cage, flares out at the two ends to contain the wire planes, LAAPD planes and their wiring.    The vessel, built out of the same low background copper used for other components and the cryostat, was designed to be very thin and hence light, owing to its proximity to the fiducial volume.   In most regions the vessel is only 1.37~mm thick, resulting in a total copper mass of $<30$~kg.   Stiffening ribs are provided for structural reasons.   While only one vessel was completed and hence no destructive test was made, extensive finite element analysis, validated by small deflection measurements on some components, found that the vessel should be capable of withstanding explosive or implosive pressures of 33~kPa with a comfortable safety factor.  Because of the many welds used in the construction, the copper was assumed to be mechanically equivalent to the weakest high purity annealed copper listed in the ASME Code (C10200 SB-187 copper rod), with minimum specified yield strength of 42~MPa and minimum specified tensile strength of 150~MPa.   Buoyancy and liquid head effects were considered in the calculations. All realistic combinations of full and empty vessel and cryostat were examined.  

\begin{figure}
	\centering
	\includegraphics[width=4in]{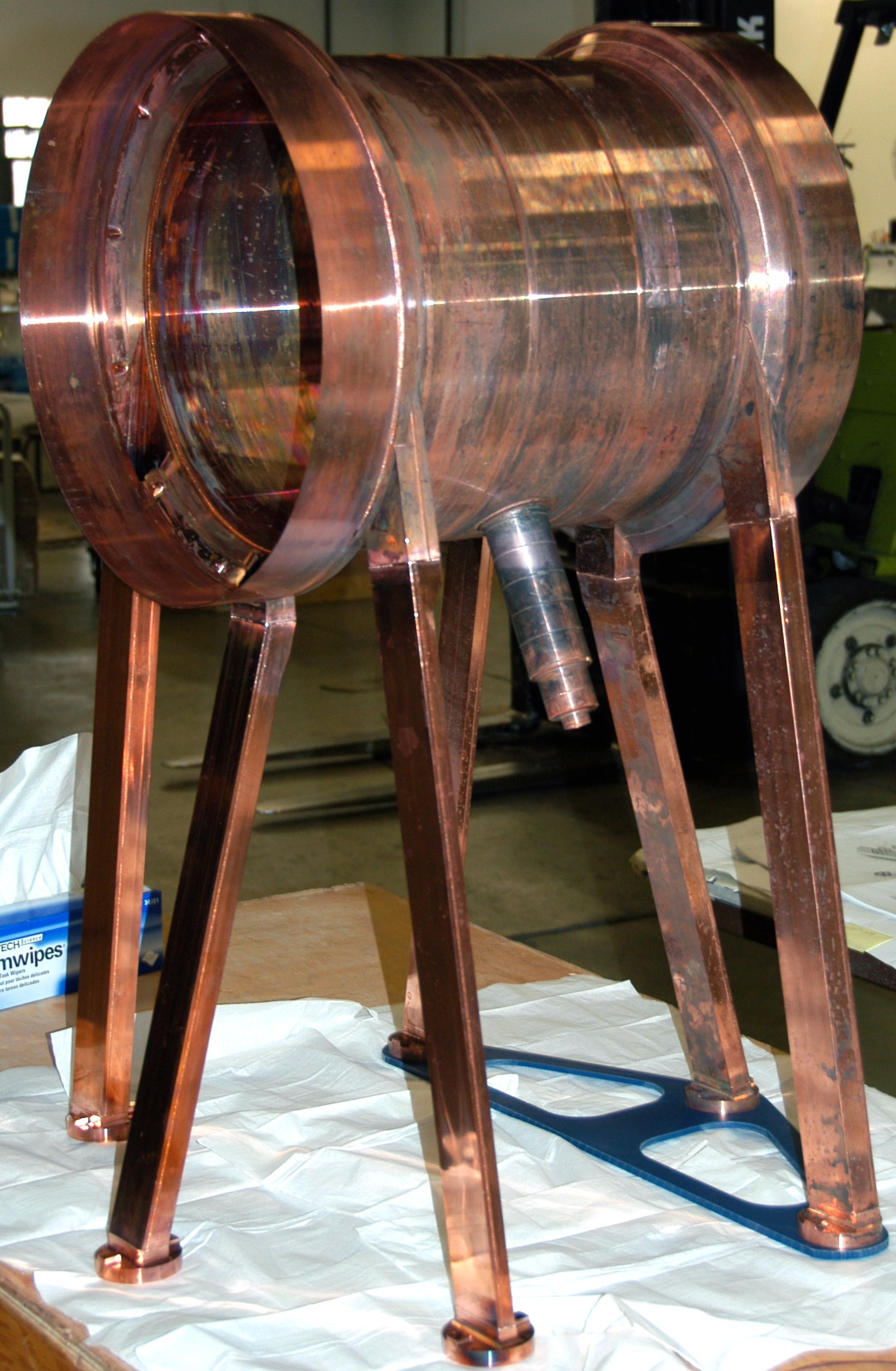}
	\caption{The completed LXe vessel before etching and welding to the cold hatch of the cryostat.
	\label{fig:lxevessel}}
\end{figure}

All copper parts are made from 5~mm or 26~mm low background stock material. All welds are of the electron-beam type, except for the ones connecting the legs to the inner cryostat hatch and final field welds to close the two bulkheads. The central cylindrical part of the vessel was obtained by rolling a 5~mm plate, welding it and then turning both inside and outside to obtain the nominal 1.37~mm thickness (except for two stiffening hoops and a socket feature at each end).   The inner diameter of this cylindrical section is 39.62~cm.  The bore for the HV feedthrough was then added.   Each of the flared regions was obtained by machining two conical sections from 26~mm stock and welding them together.    Weld preparation features were provided to mate the cylindrical section and the legs.  Special features to anchor the field cages and the wire planes and LAAPD platters were also provided in the flared sections.    The six legs (with three different cross sections for the different cable bundles and, in two cases, the pump out access and Xe recirculation) were also made by carving 26~mm and 5~mm stock and then welding them in a clam-shell configuration.   Finally, larger cylinders of 45.47~mm inner diameter were rolled and machined, using the same technique as the central cylinder, to provide weld lips for the final field weld.   Substantial fixtures were built to hold different parts together during the electron-beam welding process.    The completed vessel is shown in Figure~\ref{fig:lxevessel}.

Before inserting the TPC components, the vessel was welded to the cold hatch of the cryostat along with the tube sections comprising the HV feedthrough using a TIG welding technique (see Figure~\ref{fig:allcables}).   The two bulkheads were carved out of 26~mm stock.   Star-shaped reinforcements were then electron-beam welded on, along with a cylindrical part, designed to mate, from the inside, with the ones mounted on each end of the vessel.  Once installed, the inner surfaces of the end caps are separated by 44.5~cm.

\begin{figure}
	\centering
	\includegraphics[width=6in]{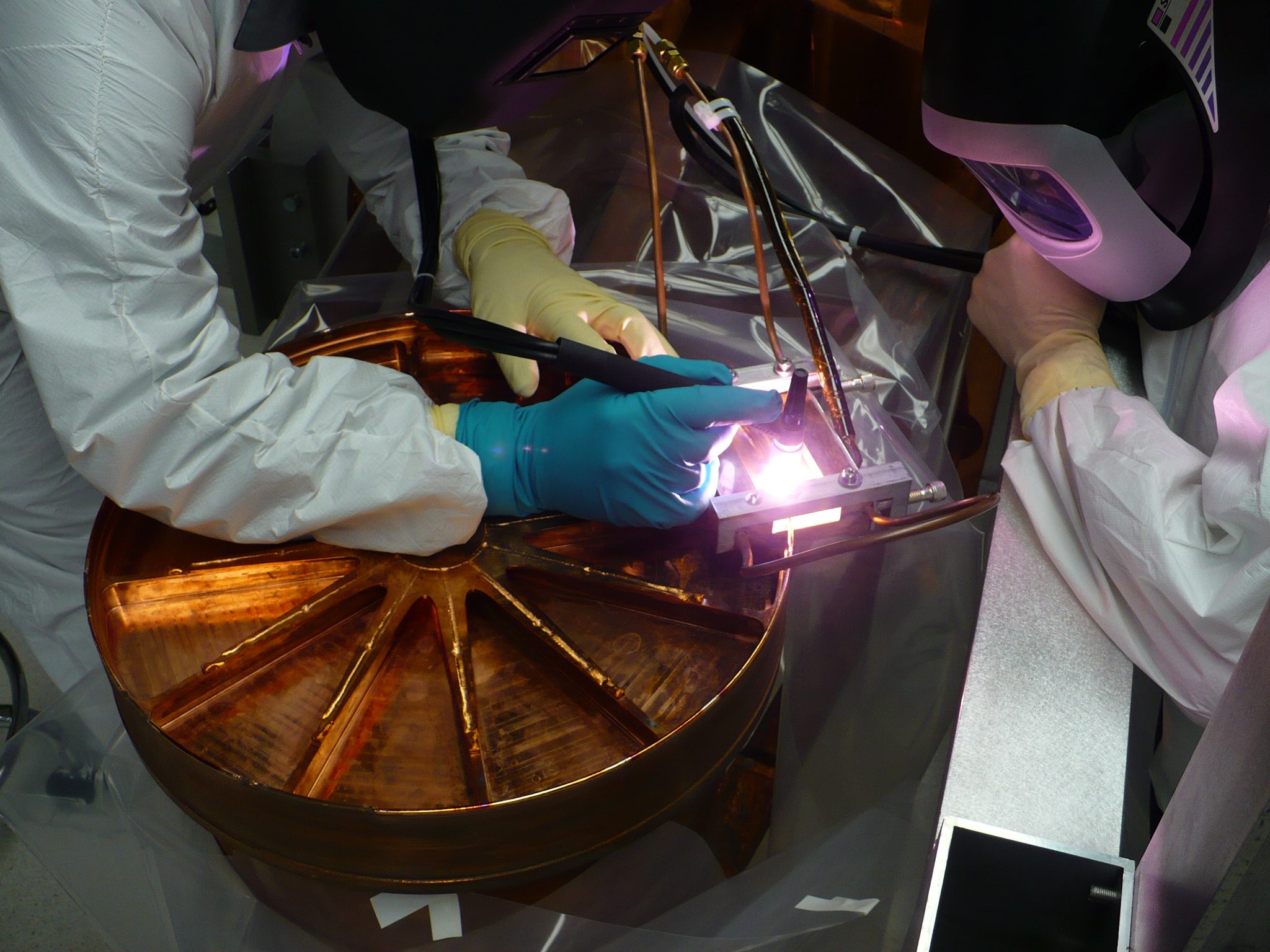}
	\caption{One the two bulkheads being TIG welded to the TPC vessel body.  The weld is performed piecemeal in the region cooled by the clamp at the 2~o'clock position.
	\label{fig:TIG_weld}}
\end{figure}

The LXe vessel is constructed from high purity electrolytic copper cast at Norddeutsche Affinerie (now Aurubis) in Hamburg Germany~\cite{aurubis}.    Copper for the vessel was in large part machined using a computer numerical control mill and a manual lathe, both sited under a shallow overburden of $\sim7$~m water equivalent. Only new carbide tools and ValCool VP-700-005-B~\cite{Valenite} coolant diluted with deionized water were used.   All copper parts were degreased and etched between processing steps and at the end, before inserting the TPC components.

Construction of the copper vessel required 47 welds, for a total of 2151~cm of welds.   The great majority (2058~cm) were electron beam welds made at Applied Fusion~\cite{aft}. In total, the welding at Applied Fusion took less than two days at sea level.    The remaining 93~cm (final two bulkhead field welds, and the connection of the HV tube to the HV feedthrough on the barrel) were TIG welded in a class 1000 cleanroom environment in the shielded building. Ceriated (as opposed to thoriated) tungsten tips were used for TIG welding to avoid transfer of radioactivity to the copper.  Specially designed cooling clamps were used to limit the temperature of the copper surrounding the welds.   This was particularly important for the final welds, sealing the detector after all components had been inserted.   As illustrated in Figure~\ref{fig:TIG_weld}, this weld was done piece-meal, while moving the cooling clamp around the circumference.    The two long weld lips designed for this seal are designed to allow cutting and re-welding of each bulkhead three times.    The use of a weld for the final seal of the detector eliminates the need for low-background, cryogenic gaskets and has proven to be a reliable solution.  Special provisions were made to He-leak check the various welds.

\subsection{Calibration source guide tube}
\begin{figure}
	\centering
	\includegraphics[width=5in]{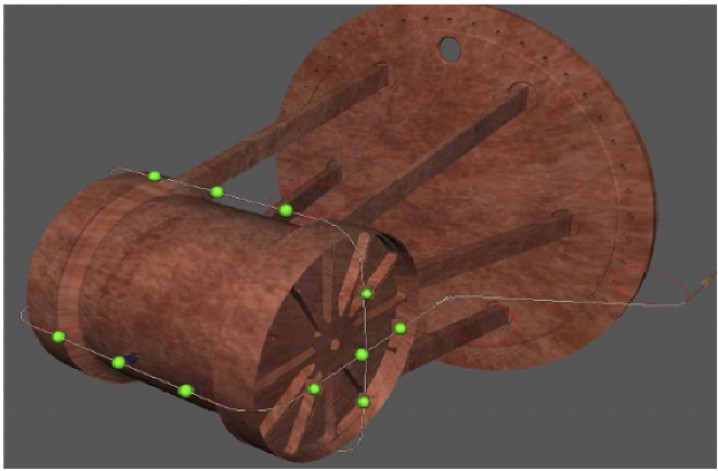}
	\caption{Calibration guide tube position around the LXe vessel.
	\label{fig:caltube}}
\end{figure}

In order to understand the response of the TPC to ionizing radiation, calibrations with radioactive \gr\ sources ($^{137}$Cs, $^{60}$Co, and $^{228}$Th) need to be performed.  The sources must be placed close enough to the detector to have large full absorption efficiencies in the active detector volume.  This is achieved by inserting sources in a  copper guide tube that wraps around the outside of the LXe vessel in the HFE-7000 volume (see Figure~\ref{fig:caltube}).  The sources can then be deployed to a known location along the length of this tube in order to access various external points around the detector.

\section{Background control}

The criteria used for the design and assembly of EXO-200 were largely dictated by the need to control backgrounds due to various types of radioactivity.   The primary methods for controlling backgrounds consist of carefully selecting construction materials, reducing surface contamination by special handling and treatment, providing adequate shielding against external radioactivity, and limiting cosmic ray activation during construction and transportation of critical components.

\subsection{Bulk contamination and material screening}
Progressively cleaner materials were chosen for the construction of detector components closer to the LXe volume and, in general, all components inside of the main 25~cm thick lead shielding were screened for radioactivity.  To simplify this task, low background detector components were designed to be made from a small number of materials: copper, phosphor bronze and silicon bronze for electrically conducting parts, acrylic (obtained from the SNO Collaboration), PTFE and polyimide substrates for dielectrics, in addition to the LAAPDs.  Additional materials were used in carefully controlled small quantities, including silicone vacuum grease (outside the LXe volume), platinum and indium plating and aluminum, MgF$_2$, and gold coatings.  Silica optical fibers were used to carry light signals directly into the TPC.  The high voltage cable was made with an extruded polyethylene dielectric.

The radioactivity screening was performed with a variety of techniques, including Neutron Activation Analysis (NAA), Gas Discharge Mass Spectrometry (GD-MS), Inductively Coupled Plasma Mass Spectrometry (ICP-MS), low background \gr\ spectroscopy, alpha counting, and $^{222}$Rn and $^{220}$Rn counting.    Over 400 materials were screened for EXO-200~\cite{Background_NIM}.

Bulk contamination in the two Xe supplies (natural and enriched) was not measured beforehand because of the difficulty of reaching the required sensitivity.   However, Xe can be purified \emph{in situ} using two SAES MonoTorr hot zirconium getters~\cite{SAES}, which are expected to remove in one pass electronegative impurities at or below the ppb level.  Kr contamination was measured by mass spectroscopy during the detector filling and, as expected, was found to be substantial in the natural Xe and drastically reduced in the $^{\textnormal enr}$Xe, as reported in~\cite{Dobi:2011zx}.  The corresponding measurements of $^{85}$Kr from the early EXO-200 data confirm such results, assuming the average concentration of $^{85}$Kr in atmospheric Kr.    Radon emanation from various components was screened with a dedicated array of seven ultra-low background electrostatic counters~\cite{andersen,wang,Jacques_detector}.   All PTFE installed in the detector volume was baked under a nitrogen purge to drive out volatile impurities.  All detector components wetted by the LXe were stored under dry nitrogen for more than 6~months before the detector was filled with LXe.  This is believed to be important to achieve good electron drift lifetimes in a short time after detector cool down.

\subsection{Surface contamination and cleaning}

Special care was taken throughout the detector assembly process to protect detector materials from surface contamination.  The detector was assembled in a class 1000 cleanroom under a shallow ($\sim$7~m water equivalent) concrete overburden.   All machining of low-background components was done with new carbide tools and always using gloves.    The only coolants or lubricants allowed were alcohol and Valenite ValCool VP700~\cite{Valenite} that had been counted beforehand.   

Before installation, all detector components received thorough surface treatment, both to remove radioactive surface contaminants and to degrease surfaces for optimal electron lifetime.  The standard procedure consisted of the following steps:

\begin{enumerate}
\item
Degreasing rinse using acetone (small parts underwent ultrasonic cleaning while immersed for $\sim$15~min, followed by a rinse with fresh solvent). Materials incompatible with acetone did not receive this treatment.
\item
Rinse using ethanol, a less aggressive degreasing solvent (small parts underwent ultrasonic cleaning while immersed for $\sim$15~min, followed by a rinse with fresh solvent). Isopropanol was substituted for ethanol when cleaning materials incompatible with ethanol. Both ethanol and isopropanol can be purchased commercially~\cite{solvents} in a form more pure than acetone.
\item
0.5~M - 1.5~M~HNO$_3$ rinse. Copper and copper-based alloys underwent the more dilute treatment. In some cases, 1-2~M~HCl was substituted (for example, in the case of the flexible cables due to the very thin copper traces).  The HNO$_3$ and HCl used were 15.8~M and 11.7~M trace-metal grade~\cite{Fisher_HNO3_HCl}, and diluted using deionized water with measured resistivity of 18~M$\Omega\cdot$cm, produced on site. 
\item
Rinse with 18~M$\Omega\cdot$cm deionized water.
\item
From the time of final cleaning, all TPC components were stored under nitrogen boil-off atmosphere.
\end{enumerate}

The efficacy of the cleaning procedure for copper was measured directly by wipe testing.  Whatman Grade-42 filter papers~\cite{whatman} were prepared by soaking for 12 to 24~hr in 2~M~HNO$_3$ to reduce their Th and U content.  They were then used to soak up 0.1~M~HNO$_3$ spread over $\sim$45~cm$^2$ of treated copper. ICP-MS analysis of these filter papers revealed a surface contamination limit of $< 2$~pg/cm$^2$ for both U and Th.

The surface cleaning procedure was modified for several components:
\begin{itemize}
\item HCl, which does not react with copper, replaced HNO$_3$ for the treatment of the fully machined and welded TPC because of the concern that HNO$_3$ would be difficult to rinse from small cavities near the many welds and would eventually damage the vessel.  Because of the difficulty of actively drying the finished vessel a final ethanol rinse was added to displace the rinse water which, in the long term, could have conceivably oxidized and corroded the copper.
\item Many components of the TPC were commercially photoetched from phosphor bronze foil. The photoetching process introduced contaminants deep into the surface of the final parts. A modified cleaning procedure was developed for these parts, consisting of a rinse in methanol, followed by a deionized water rinse, then three consecutive 10~min soaks in 3~M~HNO$_3$, each followed by a deionized water rinse. A 15$\%$ loss of sample mass was observed in using this procedure. Several photoetched components were platinum plated for improved electrical contact. Once plated, the standard cleaning procedure was applied before installation into the TPC. Those components that were not platinum plated received one final rinse in 1~M~HCl, followed by a rinse in deionized water and a rinse in ethanol just prior to installation.
\item HCl replaced HNO$_3$ when cleaning the flexible cables used to carry signals out of the TPC. A final rinse in ethanol was added to minimize copper oxidation after the deionized water rinse.
\item LAAPDs were left uncleaned from the manufacturer. Intrinsic radioactivity measurements were also done, consistently, on uncleaned devices.
\item Acrylic parts were machined ``dry'' using a special air-cooling device and received the standard procedure substituting a 25$\%$ Isopropanol solution for Acetone and Ethanol, and 1~M~HCl for HNO$_3$.
\item The outer surface of the cryostat was cleaned using the standard cleaning procedure. The inner surfaces of the cryostat were cleaned using the standard procedure during fabrication, and once received were cleaned again inside a class 1000 cleanroom using only steps 1-2 of the standard procedure (because of concerns that a complete acid removal from welds was difficult to achieve).
\item Superinsulation from Sheldahl \cite{sheldahl} with single sided aluminization is used between the walls of the cryostat.  The superinsulation is extremely fragile, and so was not cleaned once received from the manufacturer. Intrinsic radioactivity measurements were performed without prior cleaning and indicated an acceptable cleanliness. The superinsulation was handled only in a class 1000 cleanroom.
\item Lead shielding bricks were painted with an epoxy and treated with steps 1-2 of the standard cleaning procedure only.
\end{itemize}

\subsection{Passive shielding and Rn enclosure}

Nested layers of passive shielding isolate the LXe from external radioactivity in EXO-200. Shielding layers are selected to be progressively cleaner as they get closer to the active Xe volume.   The innermost shielding is provided by a $\ge 50$~cm thick layer of HFE-7000~\cite{HFE} contained in the inner vessel of the cryostat.  The two nested vessels of the cryostat provide further shielding of  5.4~cm of copper.   Finally, the outermost shielding layer is made out of 25~cm of interlocking low radioactivity lead blocks.   The radioactivity content of all shielding materials is given in~\cite{Background_NIM}.

All services enter the cryostat from the front hatches, through penetrations in the primary lead shielding wall.    In order to mask the direct line of sight, the services undergo a 90$^{\circ}$ bend immediately outside of such a wall.   A secondary 20~cm thick lead wall is then built.    Services are made out of low activity materials up to the region past the 90$^{\circ}$ bend, beyond which conventional materials (mainly stainless steel) are used.   The front-end electronics, built out of conventional (non low-background) components, are mounted between the two lead walls and out of the direct line of site of the TPC. The general shielding layout of EXO-200 is shown in Figure~\ref{fig:cleanroom}.

The lead for the EXO-200 shielding was purchased from the Doe Run Company~\cite{doerun} and cast and machined by JL~Goslar~\cite{goslar}.    Different shapes and sizes of blocks were used for different regions of the shielding.  Gaps between blocks measured $<$1~mm, with no direct line of sight through the shield once both front walls are installed. A 10~$\mu$m clear epoxy coating encapsulates each brick, maintaining clean and oxide-free surfaces.  

A 0.8~mm thick stainless steel sheet metal enclosure surrounds the lead shielding for future use to displace the radon from the volume of air inside the lead shield.

\subsection{Cosmic-ray veto system}
The vertical cosmic-ray muon flux has been measured to be $(3.10 \pm 0.07)\times10^{-7}\,\textnormal{s}^{-1}\textnormal{cm}^{-2}\textnormal{sr}^{-1}$ at WIPP~\cite{muonflux}.  Muons traversing the TPC are easily rejected as a background by their large energy deposition and track-like signals.    However, an external veto system is required to reject muon bremsstrahlung and \gr s emitted by the prompt relaxation of nuclear excitations; the latter being induced in the detector components and shielding by passing cosmic-ray muons and spallation neutrons.  This was achieved by an array of plastic scintillator panels externally installed on four of the six sides of the clean room module containing the TPC, as shown in Figure~\ref{fig:cleanroom}. Thirty-one 5~cm thick Bicron BC-412 plastic scintillator panels were obtained from the KARMEN neutrino oscillation experiment~\cite{KARMEN} for this purpose.  Eleven of the panels are 375~cm long by 65~cm wide; twenty are 315~cm long by 65~cm wide. The panels are equipped with 180\deg\ light guides at each end.  Each panel end is read out by four 2'' Photomultiplier tubes (PMTs) glued with optical cement to the light guides.   Panels are wrapped in crinkled aluminum foil to increase the light collection. The underside of each panel is covered by 4~cm of borated polyethylene (5\% loading by mass).  The polyethylene not only provides structural support to the panel but acts as a partial thermal neutron shield for the TPC.  Before installation at WIPP, all panels were refurbished, tested, and calibrated.  A total of 29 panels were installed. 

A total of 232 Philips model XP2262 PMTs are used in the veto detector, gain matched in groups of four.   PMTs on each panel end are powered by a single HV channel and readout together.   Voltages between 1.4 and 1.7~kV are provided by a CAEN Model Sy527 universal multichannel power supply system, resulting in an average PMT gain of $4.96\times 10^6$, with a standard deviation of 2.6\%.  A dedicated readout module accepts analog signals from the 58 channels, applies a threshold and reports the pattern of channels above threshold for inclusion in the TPC data stream.   

The efficiency of the veto detector for muons traversing the TPC was projected to be greater than 95\% using simulation and measured, from early EXO-200 data, to be ($96.0\pm0.5$)\%.  The main source of inefficiency is the incomplete coverage of the clean room by the scintillator panels.

\subsection{Activation control and installation}

Cosmic-ray activation of various detector components is a common concern in low-background experiments.   Given the large amount of copper in EXO-200, the production of $^{60}$Co is particularly insidious.  The copper cryostat and all copper TPC components were made from two production batches of high purity electrolytic copper~\cite{aurubis} rolled into 26~mm and 5~mm plates.   The first batch of copper was used to make the cryostat, containing $\sim$2700~kg of copper.  The cryostat was machined in Grenoble, France~\cite{sdms} with components stored in a shallow underground site during production.   The cryostat and remaining copper were then shipped to the United States by surface and stored for over a year under a $\sim$7~m water equivalent concrete overburden.    After installation in the clean room and initial testing, the entire setup was shipped to WIPP by road in 2008 (about 24~hrs at sea level) and then stored underground for three years before the beginning of the low background data taking.  

Most TPC and copper vessel components were machined from the first batch of copper.  A second batch of copper was produced and used to machine several components of the copper vessel.  After casting, the copper ingot was stored for 90 days in a concrete bunker at DESY~\cite{desy}, waiting for rolling into the 26~mm and 5~mm plates.    The copper then spent 20 days at sea level during the rolling process and was transported to the United States by sea in a 2~m water equivalent shielded shipping container, taking 45~days.    The machining of all copper components of the TPC was performed under 7~m water equivalent overburden in a dedicated machine shop.  The only exception was the electron-beam welding process for which the components spent an integrated time of $<2$~days at sea level.   

The TPC, including the copper vessel welded to the inner cryostat hatch, was shipped to WIPP in November 2009.  In preparation for transportation to WIPP, the TPC was purged with a 14~kPa overpressure of boil-off nitrogen.  It was then sealed in three layers of 150~$\mu$m thick class 100 polyethylene film and enclosed in an aluminum sheet metal box for protection.   To protect the TPC from shocks and vibrations during the 2100~km road trip from Stanford University to WIPP, the TPC was mounted to a specially designed vibration-damping pallet.  

\begin{figure}
	\centering
	\includegraphics[width=6in]{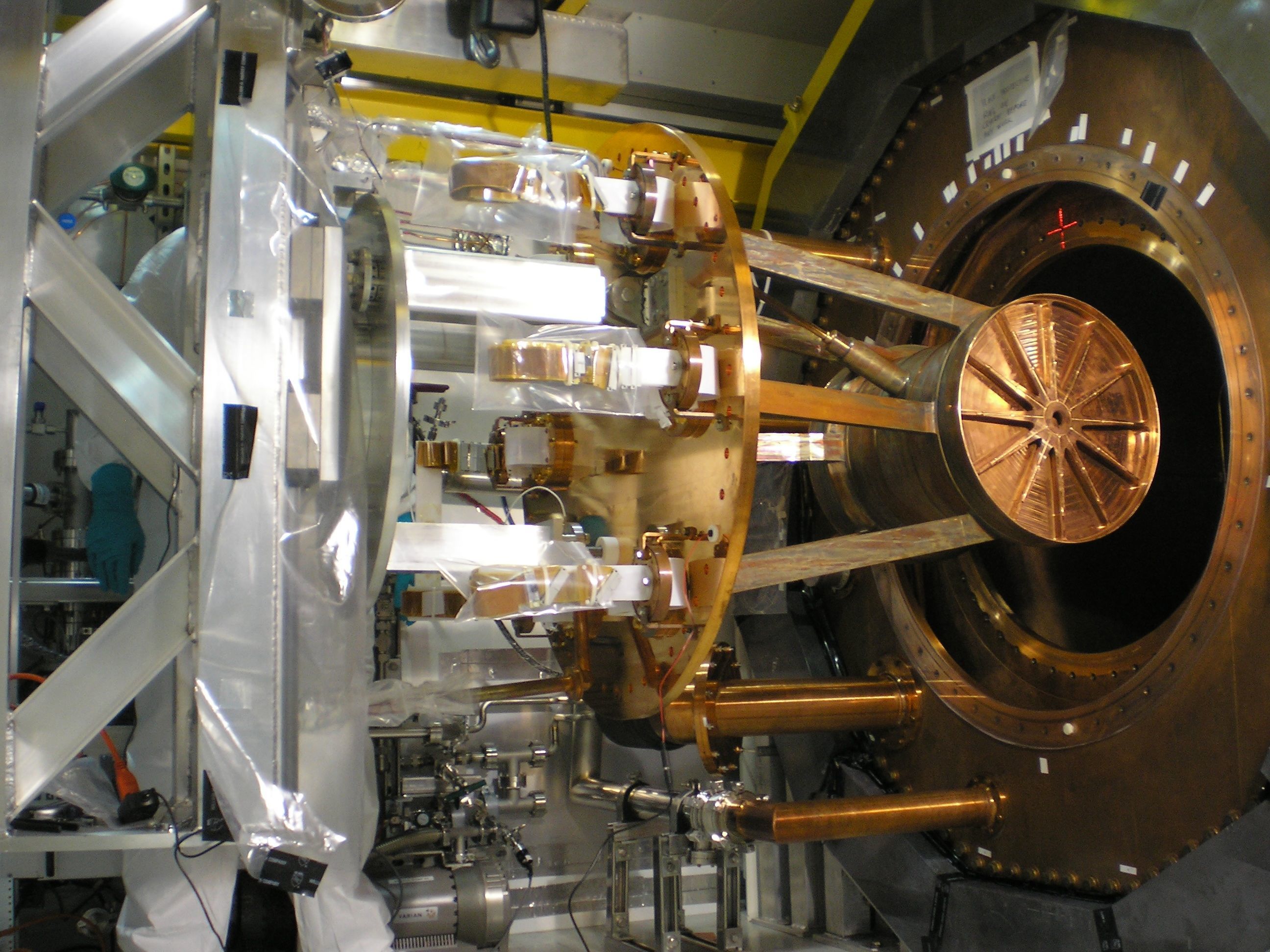}
	\caption{The EXO-200 TPC on the installation jig, being inserted in the cryostat.
	\label{fig:TPC_installation}}
\end{figure} 

Cosmic ray activation of detector components during shipping was minimized by optimizing the driving route and housing the TPC in the same shielded container built for the copper transportation.   The concrete thickness was optimized to saturate the maximum allowed load on a standard non-escorted vehicle ($\sim40$~tons).    According to FLUKA Monte Carlo simulations~\cite{FLUKA} and tabulated cross sections, the vault reduced the $^{60}$Co production from cosmic radiation by a factor of $3.7\pm0.24$, as discussed in more detail below.  The TPC was underground for 18 months before the start of the low background data taking.

The TPC was installed in the cryostat in January 2010.  The installation was done using a custom machine with three translational and three rotational degrees of freedom, as shown in Figure~\ref{fig:TPC_installation}.  Load cells were used to measure the weight and pitch of the detector-hatch system throughout the installation process.  Laser cross-hair generators were used for alignment of the hatch with the cryostat bolt holes.

\section{Background modeling}

A detector Monte Carlo model based on GEANT 3.21~\cite{GEANT321} was developed to inform the material selection and detector design. This was done with the goal of creating a simulation tool flexible enough to follow frequent early design changes and select materials on the basis of their background contribution.

Complex decay sequences were modeled by custom-coded event generators in order to include particle correlations. All decays were modeled as fully stochastic in terms of their sequence, energy, and particle content. $\beta$ decays were modeled using realistic spectral parameterizations.   Spectral corrections for unique forbidden decays were implemented while non-unique decays were modeled as allowed.   The detector response was modeled schematically, folding the simulated energy deposits in the LXe with Gaussian response functions in both energy ($\sigma / E = 1.5$\% at the double beta decay endpoint) and space ($\sigma = 1$~cm).

A simplified event reconstruction was setup to form event vertexes as the energy-deposit weighted sum over all interaction points.  Event multiplicity cuts were performed on a phenomenological track length variable defined, separately, for each coordinate, as the largest distance between any points along a track.  This is basically the smallest coordinate box that can contain the event. Analysis tracking cuts were made in each of the three variables independently.  Multi-site events were identified by track lengths in excess of 2~cm.

For each candidate material the radioactivity was measured and its impact on the detector design estimated using the simulation.   As the detector design and construction spanned a period of several years, the final estimate of the full background was only available at the end of the process. In order to provide guidance along the design and selection process the following criteria were generally enforced: large components, external to the TPC and forming significant shielding layers, were allowed to contribute no more than 10\% of the allowable total zero neutrino background rate; internal TPC components (of which there are many) were allotted a 1\% background contribution; a 1\% tolerance was also applied to small external parts.  In general these restrictions were strictly applied to generic parts (seeking different stock if needed) but somewhat relaxed for a few essential components that would have been too hard to replace.   The overall background allowance was set both in terms of impact of the \zeronubb\ and \twonubb.   The background in an energy interval of $\qbb\pm2\sigma(\qbb)$ (the \zeronubb\ analysis window) was limited to 33~events/yr in a 110~kg active Xe mass, the expected event rate for \xeonethirtysix\ consistent with the claim by~\cite{HVKK} in their early publications.    The background in the \twonubb\ analysis region (400--2000~keV) was limited to 26400 events per year, consistent with the upper limit~\cite{Bernabei} available at the time of construction.  In practice the requirements for the \zeronubb\ tend to constrain the Th and U content of materials, while the \twonubb\ provides looser constraints but is affected by $^{40}$K contamination.

\subsection{\gr\ leakage through passive shield}

A total of 23 salt samples were collected at WIPP and subsequently counted, using a shielded Ge detector, to establish the activity environment in the detector area. Table~\ref{tab:salt} lists the average specific activities with standard deviations dominated by the sample to sample differences.

To calculate the salt-related background by Monte Carlo, a boosting scheme was developed in order to reduce the computational time to a manageable level. To decouple the solid angle calculation from the radiation transport through the passive lead and copper shielding, 2.6~MeV $\gamma$-radiation was tracked, from its point of origin somewhere in the salt,  until it reached any of the outer surfaces of the lead shield. The energy and direction of the \gr\ were then histogrammed and the hit efficiency recorded. The radiation transport through the shield was simulated after accumulating sufficient statistics to produce a full \gr\ flux field at the outer surface of the lead shielding.  This flux field was then treated as a probability distribution from which to sample background radiation at the surface of the lead shield.    Since this procedure neglects energy-direction correlations its validity was verified by a comparison with the full simulation for a reduced, 10~cm thick lead shield.  The two methods agreed to within 30\%, and the boosting scheme was found to increase the LXe hit efficiency by a factor of 400. 

Isotropic 2.615~MeV $\gamma$-radiation was simulated with starting points homogeneously distributed within a 1~m thick salt layer around the experimental hall. This thickness is essentially infinite as 87\% of the photons that hit the detector were found to originate within the first 20~cm of salt.  The thickness of the lead shield was set by the background requirements discussed above.  The projected backgrounds from the salt are also listed in Table~\ref{tab:salt}.    Lower energy \gr s from the salt were found to give negligible contributions to the background.  

\begin{table}
  \centering
    \begin{tabular}{c c c c c c}
      \lighttoprule\lighttoprule
      \multicolumn{3}{c}{Average specific Activity [Bq/kg]} & $\phantom{0}$ & \twonubb\ background & \zeronubb\ background \\
      \cmidrule{1-3}
      \kforty   & \thtwothirtytwo & \utwothirtyeight && [counts/yr] & [counts/yr]\\
      \midrule
      $60\pm47$ & $0.54\pm 0.42$ & $0.76\pm 0.69$ && $548\pm11$ & $1.3^{+1.4}_{-0.9}$ \\
      \bottomrule
    \end{tabular}
  \caption{\label{tab:salt}Average specific activities found in 23 salt samples taken from WIPP and their impact on the EXO-200 background.  The error quoted on the specific activity reflects standard deviation in sample to sample variations.}
\end{table}

\subsection{Backgrounds from the detector components}
The activity of various detector components, some of which are reported in~\cite{Background_NIM}, along with their masses, were used as inputs for the simulation.    Decay generators were developed to model all of the branches of the \utwothirtyeight\ and \thtwothirtytwo\ decay series (with $>1\%$ branching ratio) and the beta decay of \kforty.  The full decay generators were employed to simulate natural radioactivity in the materials installed within the TPC and the TPC itself.  To reduce computation time, only the \gr s were modeled for materials located outside of the TPC.   In all but one case (\pbtwoten, see below), the \utwothirtyeight\ and \thtwothirtytwo\ decay series were assumed to be in secular equilibrium.  Table~\ref{tab:naturalradioactivity} shows the activities, quantities, and impact on the \zeronubb\ and \twonubb\ of detector components with largest impact.

A special case is \pbtwoten\ that is accumulated in the massive lead shield.   Its concentration in the EXO-200 lead was determined by $\alpha$ counting to be 30~Bq/kg for a total activity of roughly 2~MBq.    A truncated model of the \utwothirtyeight\ series, starting with \pbtwoten, was used to look for bremsstrahlung and a low probability 803~keV \gr\ produced by the subsequent decays of \bitwoten\ and \potwoten. This results in $<800$~events/yr in the \twonubb\ analysis region, with slightly more stemming directly from the \potwoten\ \gr.  Due to the low energies involved, these decays do not contribute to the \zeronubb\ analysis region.

\begin{sidewaystable}
  \scriptsize
  \centering
  \begin{tabular}{l c c c c c c c c c c c}
    \lighttoprule\lighttoprule
    &&\multicolumn{3}{c}{Radioactivity [mBq]} & $\phantom{0}$ & \multicolumn{3}{c}{ 2$\nu$ background [counts/yr]} & $\phantom{0}$ & \multicolumn{2}{c}{ 0$\nu$ background [counts/yr]}\\
    \cmidrule{3-5} \cmidrule{7-9} \cmidrule{11-12}
    Part/material & Quantity & K & Th & U && K & Th & U && Th & U\\
    \midrule
    APDs                & \phantom{}518 units     & $\phantom{0000}<0.13\phantom{0}$    & $\phantom{000}<0.09\phantom{00}$ & $\phantom{000}<0.011\phantom{0}$ && $\phantom{000}<39\phantom{.00}$ & $\phantom{0}<310\phantom{.0}$ & $\phantom{0}<340\phantom{.0}$ && $\phantom{}<1.0\phantom{000}$ & $\phantom{0}<1.5\phantom{00}$\\
    Bronze cathode         & \phantom{0}0.010\,kg    & $\phantom{0000}<0.019\phantom{}$    & $0.00108\phantom{}\pm0.00019\phantom{}$ & $0.00364\phantom{}\pm0.00021\phantom{}$ && $\phantom{00}<340\phantom{.00}$ & $28\pm\phantom{0}5$ & $\phantom{}193\pm\phantom{}11$ && $0.0071\phantom{}\pm0.0012\phantom{}$ & $1.11\phantom{0}\pm0.06\phantom{0}$\\
    Bronze wires           & \phantom{0}0.083\,kg    & $\phantom{0000}<0.16\phantom{0}$    & $0.0090\phantom{0}\pm0.0015\phantom{0}$ & $0.0302\phantom{0}\pm0.0017\phantom{0}$ && $\phantom{000}<50\phantom{.00}$ & $32\pm\phantom{0}6$ & $\phantom{0}84\pm\phantom{0}5$ && $0.110\phantom{0}\pm0.019\phantom{0}$ & $0.370\phantom{}\pm0.021\phantom{}$\\
    Other bronze  & \phantom{0}0.314\,kg    & $\phantom{0000}<0.6\phantom{00}$    & $0.0176\phantom{0}\pm0.0027\phantom{0}$ & $0.107\phantom{00}\pm0.006\phantom{00}$ && $\phantom{00}<120\phantom{.00}$ & $63\pm\phantom{}10$ & $\phantom{}295\pm\phantom{}17$ && $0.216\phantom{0}\pm0.033\phantom{0}$ & $1.30\phantom{0}\pm0.08\phantom{0}$\\
    Flat cables\\
    \quad in TPC        & \phantom{0}7406\,cm$^2$ & $\phantom{0000}<0.9^{\dagger}\phantom{0}$    & $\phantom{000}<0.07\phantom{00}$ & $0.43\phantom{}\pm0.06\phantom{}$ && $\phantom{00}<250\phantom{.00}$ & $\phantom{0}<220\phantom{.0}$ & $\phantom{}1080\pm\phantom{}160$ && $\phantom{}<0.8\phantom{000}$ & $4.6\phantom{00}\pm0.7\phantom{00}$\\
    \quad in TPC legs   & \phantom{}10825\,cm$^2$ & $\phantom{0000}<1.4^{\dagger}\phantom{0}$    & $\phantom{000}<0.07\phantom{00}$ & $0.76\phantom{}\pm0.09\phantom{}$ && $\phantom{000}<19\phantom{.00}$ & $\phantom{00}<10\phantom{.0}$ & $\phantom{00}76\pm\phantom{00}9$ && $\phantom{}<0.07\phantom{00}$ & $0.262\phantom{}\pm0.032\phantom{}$\\
    Teflon reflectors   & \phantom{0}1.530\,kg    & $0.087\phantom{}\pm0.010\phantom{}$ & $\phantom{000}<0.0022\phantom{}$ & $\phantom{000}<0.008\phantom{0}$ && $40\pm4$ & $\phantom{00}<12\phantom{.0}$ & $\phantom{00}<60\phantom{.0}$ && $\phantom{}<0.034\phantom{0}$ & $\phantom{0}<0.32\phantom{0}$\\
    Teflon behind APDs  & \phantom{0}3.375\,kg    & $0.193\phantom{}\pm0.021\phantom{}$ & $\phantom{000}<0.005\phantom{0}$ & $\phantom{000}<0.017\phantom{0}$ && $51\pm6$ & $\phantom{00}<15\phantom{.0}$ & $\phantom{00}<82\phantom{.0}$ && $\phantom{}<0.06\phantom{00}$ & $\phantom{0}<0.35\phantom{0}$\\
    Acrylic spacers\\
    \quad and insulators& \phantom{0}1.460\,kg    & $\phantom{0000}<0.14\phantom{0}$    & $\phantom{000}<0.024\phantom{0}$ & $\phantom{000}<0.07\phantom{00}$ && $\phantom{000}<65\phantom{.00}$ & $\phantom{0}<130\phantom{.0}$ & $\phantom{0}<290\phantom{.0}$ && $\phantom{}<0.37\phantom{00}$ & $\phantom{0}<1.6\phantom{00}$\\
    Field cage resistors& \phantom{0}20 units     & $\phantom{0000}<0.08\phantom{0}$    & $\phantom{000}<0.0006\phantom{}$ & $\phantom{000}<0.0017\phantom{}$ && $\phantom{000}<35\phantom{.00}$ & $\phantom{000}<3.0\phantom{}$ & $\phantom{000}<0.5\phantom{}$ && $\phantom{}<0.009\phantom{0}$ & $\phantom{0}<0.038\phantom{}$\\
    Cu TPC              & \phantom{}32.736\,kg    & $\phantom{000}<60\phantom{.000}$    & $\phantom{000}<0.5\phantom{000}$ & $\phantom{000}<1.5\phantom{000}$ && $\phantom{}<13000\phantom{.00}$ & $\phantom{}<1100\phantom{.0}$ & $\phantom{}<2600\phantom{.0}$ && $\phantom{}<5\phantom{.0000}$ & $\phantom{}<12\phantom{.000}$\\
    Cu TPC legs         & \phantom{0}6.944\,kg    & $\phantom{000}<12\phantom{.000}$    & $\phantom{000}<0.11\phantom{00}$ & $\phantom{000}<0.33\phantom{00}$ && $\phantom{00}<170\phantom{.00}$ & $\phantom{00}<13\phantom{.0}$ & $\phantom{00}<33\phantom{.0}$ && $\phantom{}<0.10\phantom{00}$ & $\phantom{0}<0.11\phantom{0}$\\
    HV cable            & \phantom{0}0.091\,kg    & $\phantom{0000}<5\phantom{.000}$    & $\phantom{000}<0.036\phantom{0}$ & $\phantom{000}<0.6\phantom{000}$ && $\phantom{00}<160\phantom{.00}$ & $\phantom{00}<12\phantom{.0}$ & $\phantom{0}<150\phantom{.0}$ && $\phantom{}<0.07\phantom{00}$ & $\phantom{0}<0.6\phantom{00}$\\
    Cu calibration\\
    \quad tube          & \phantom{0}0.473\,kg    & $\phantom{0000}<8\phantom{.000}$    & $0.016\pm0.003$                  & $0.043\phantom{}\pm0.001\phantom{}$ && $\phantom{0}<1100\phantom{.00}$ & $18.5\pm3.8\phantom{0}$        & $45\pm12$ && $0.100\pm0.021$ & $0.18\pm0.05$\\
    Cu wire calibration\\
    \quad
    tube support        & \phantom{0}0.144\,kg    & $\phantom{000}<11\phantom{.000}$    & $0.027\pm0.002$                  & $0.19\phantom{}\pm0.06\phantom{}$ && $\phantom{00}<700\phantom{.00}$ & $16.7\pm1.2\phantom{0}$        & $10\pm\phantom{0}3$ && $0.097\pm0.007$ & $0.04\pm0.01$\\
    Cu cryostat         & \phantom{0}5901\,kg     & $\phantom{000}<72\phantom{.000}$    & $\phantom{00}<19\phantom{.0000}$ & $\phantom{00}<58\phantom{.0000}$ && $\phantom{0000}<9\phantom{.00}$ & $\phantom{00}<29\phantom{.0}$ & $\phantom{00}<46\phantom{.0}$ && $\phantom{}<0.4\phantom{000}$ & $\phantom{0}<0.19\phantom{0}$\\
    HFE-7000\\
    \quad
    cryogenic fluid     & \phantom{0}4140\,kg     & $\phantom{000}<20\phantom{.000}$    & $\phantom{000}<0.25\phantom{00}$ & $\phantom{000}<0.8\phantom{000}$ && $\phantom{00}<220\phantom{.00}$ & $\phantom{00}<27\phantom{.0}$ & $\phantom{00}<65\phantom{.0}$ && $\phantom{}<0.20\phantom{00}$ & $\phantom{0}<0.25\phantom{0}$\\
    Pb brick paint      & \phantom{0}0.300\,kg    & $\phantom{0000}<8\phantom{.000}$    & $\phantom{000}<0.17\phantom{00}$ & $3.00\phantom{}\pm0.30\phantom{}$ && $\phantom{0000}<0.19\phantom{}$ &&&&$\phantom{}<0.0015\phantom{}$ & $\phantom{0}<0.004\phantom{}$  \\
    Pb shielding        & \phantom{}55000\,kg     & $\phantom{}<33000\phantom{.000}$    & $\phantom{}<2700\phantom{.0000}$ & $\phantom{}<8300\phantom{.0000}$ && $\phantom{000}<40\phantom{.00}$ & $\phantom{00}<60\phantom{.0}$ & $\phantom{00}<85\phantom{.0}$ && $\phantom{}<0.9\phantom{000}$ & $\phantom{0}<0.5\phantom{00}$\\
    \bottomrule
  \end{tabular}
  \caption{\label{tab:naturalradioactivity}Total radioactivity of the major components of EXO-200 and of the components internal to the TPC.  Pb shielding does not include the outer front wall.  $^{\dagger}$Because the installed cables were not assayed for K contamination, these activities are derived from assays of similar materials.}
\end{sidewaystable}

\subsection{External radon}

The background impact of radon (\rntwotwentytwo), present in the air inside the lead shielding, was also estimated with the GEANT 3.21 Monte Carlo simulation.  The air used for the decay vertex generation had a volume of 204~L, dominated by a gap between the front end of the cryostat and the lead wall.  This space is required by the routing of various feedthroughs and services to the detector.   

To decrease computation time, only the \gr\ emissions from the \utwothirtyeight\ decay series after \rntwotwentytwo\ were simulated.  An upper limit on the \rntwotwentytwo\ concentration allowable in the air inside the lead shielding of 2~mBq/L was set by requiring fewer than 0.3~counts/yr (1\% of the target rate) in the \zeronubb\ analysis region.  Measurements of radon in the clean room air indicate an average activity about three times this level which favors, for the future, the purging of the volume inside the lead shield with reduced radon air.

\subsection{Cosmogenic radioactivity}

As discussed, care was taken to prevent excessive cosmogenic activation of metals used in the detector construction and, in particular, the large copper parts. Backgrounds from the decay of \cosixty, \cofiftyeight, \cofiftysix, \mnfiftyfour, \fefiftynine, and \znsixtyfive\ were all studied.  The GEANT 3.21 simulation was used to determine the potential background impact of these radioisotopes contained in the copper parts. Only \cofiftysix\ has a single emission above the \zeronubb\ endpoint. However, \cosixty\ produces coincident \gr s with a total energy above \qbb\ and sources near the detector can result in the detection of both \gr s at the same site.   All of the isotopes discussed in this section can contribute to the \twonubb\ backgrounds.
\clearpage

Cosmogenic activation rates were calculated using production yields along with exposure models based on the material handling described in previous sections.   The surface production rate limits were measured by exposing samples of copper to cosmic radiation above ground followed by storage and counting with an underground germanium detector~\cite{weber}.  This sequence was repeated several times.  For \cosixty\ the measurement from Heusser~\cite{Heu93}, which was below our measurement limit, was used.  The production rates at the shielded assembly facility were roughly estimated by scaling the surface production rates due to the fast secondary cosmic ray neutron flux, given by Gordon~\cite{ziegler}, as a function of shielding depth from a FLUKA simulation~\cite{FLUKA}.  The scaling was cross-checked to agree reasonably well with the depth dependence extracted from Heusser~\cite{Heu95}.

A detailed geometry was simulated for the shielded truck used to transport the TPC from California to New Mexico, having a shielding depth of $\sim$2~m water equivalent with $2\pi$ coverage.  The neutron spectrum was stabilized by passing it through a thick layer of air.  Production rates for \cosixty\ were then calculated above and inside the truck using tabulated cross sections serving as a benchmark.  The shielded truck was found to reduce production rates by nearly a factor of four.  

A summary of the expected cosmogenic activity in the detector copper, along with its projected backgrounds, are shown in Table~\ref{tab:cosmogenicsactivity}.  These activities assume 60~days of cooling underground at WIPP (although a much longer time elapsed between the transportation of the TPC to WIPP and the beginning of the low background data taking).

\begin{table}
   \centering
  \begin{tabular}{l c c c c}
    \lighttoprule\lighttoprule
            &          & Projected activity & \twonubb\ background & \zeronubb\ background \\
    Isotope & Half-life & [$\mu$Bq/kg]       & [counts/yr]          & [counts/yr] \\
    \midrule
    \mnfiftyfour  & \phantom{}312\,d    &$<15$ & $<1000$ \\
    \cofiftysix   & \phantom{0}77\,d    &$<10$ & $<1200$ & $<5\phantom{.00}$\\
    \cofiftyeight & \phantom{0}71\,d    &$<27$ & $<2000$\\
    \fefiftynine  & \phantom{0}45\,d    &$<22$ & $<1600$\\
    \cosixty      & \phantom{.-.--}5\,yr&$<10$ & $<1100$ & $<0.24\phantom{}$\\
    \znsixtyfive  & \phantom{}244\,d    &$<27$ & $<1000$\\
    \bottomrule
  \end{tabular}
  \caption{\label{tab:cosmogenicsactivity}Main sources of cosmogenic activity expected to be produced based on the copper exposure described in the text.  Activities and backgrounds are calculated at the time that the detector was brought underground at WIPP.  In the table, missing numbers correspond to negligible contributions.}
\end{table}

\subsection{Muon induced background}

Muons traversing the LXe in the TPC produce unmistakably high-energy signals.  However, muons passing near the detector can produce \gr s which can then deposit lower energies in the TPC, even producing single-site double beta decay-like events. For this reason, muon backgrounds were also investigated using the GEANT 3.21 detector simulation.  Muons were generated  with a ratio of $\mu^+/\mu^-$ of 1.25.  The muon energy spectrum was parameterized in varying degrees of detail using equations from Gaisser~\cite{Gai90}. Ultimately, the following form was used:

\begin{equation}
  \frac{\text{d}N}{\text{d}E}\propto E_0^{-(\gamma+1)}\left(\frac{1}{1+1.1 E_0/(115\,\text{GeV})}+\frac{0.054}{1+1.1 E_0/(850\,\text{GeV})}\right),
\end{equation}

where $N$ is the number of muons, $\gamma=1.7$ and $E_0$ is the energy in GeV that the muon had at the surface.  This energy is related to the energy $E$ at depth via

\begin{equation}
  E_0=e^{bh} (E+\epsilon)-\epsilon.
\end{equation}

Here the depth in the experimental hall at WIPP is $h= 1.584 \times 10^5$~g/cm$^2$, $b=0.4$ and $\epsilon = 550$~GeV.    The angular distribution,

\begin{equation}
  \frac{\text{d}N}{\text{d}\theta} \propto \cos^{1.53}(\theta)e^{-8.0\times10^4 h/\cos(\theta)}
   \sin(\theta)\,,
\end{equation}

is based on Miyake~\cite{Miy73}.   $\tfrac{\text{d}N}{\text{d}\theta}$ goes to zero at $\theta=0$ as a result of the trailing solid angle factor. At the stated depth, the maximum of the angular distribution is at approximately $30^{\circ}$ and extends with significant probability above $60^{\circ}$.   Angle-energy correlations were not considered in the simulation.

Muons were generated on a horizontal plane approximately 3~m above the TPC center over an area typically about $35\,\text{m}\times 35\,\text{m}$.  This rather large area makes the computation inefficient but was found to be necessary to properly account for muon trajectories with large zenith angles, which become particularly important when comparing different veto geometries.  To improve the speed of the simulation, the predicted impact parameter of each generated muon to the center of the detector was computed, and only muons with this parameter below 1.5~m were further tracked.   To verify that this proximity cut did not introduce a bias, one simulation representing several years of muon exposure was performed with an expanded trajectory cut. No $\znbb$ candidate background events were seen for muons failing the 1.5~m cut, although a few $\tnbb$ candidate energy depositions were observed.  To normalize the simulation to a certain live time, the observed integral muon flux of $(4.09\pm0.4)\times10^{-7}$\,cm$^{-2}$s$^{-1}$~\cite{muonflux} was used.  

This process predicts $31\pm1$ muon induced \zeronubb\ background events per year if no external muon veto is employed. This rate is ten times higher than the goal of 3~events/yr.   The plastic scintillator veto already described reduces the rate of these events to negligible levels.

\section{Electronics}

\subsection{Architecture}

The front-end electronics (FEE) system is located immediately outside of the inner front lead wall and is connected to the U, V wires and LAAPDs by the thin copper-clad polyimide cables.   The signal processing architecture is identical for the three types of signals, except for the appropriate coupling capacitors and the bias voltages applied.

A FEE channel consists of a low noise charge amplifier coupled to two shaper stages, each consisting of an integrator and differentiator. This is followed by a sample-and-hold circuit and a 12~bit, 1~MS/s Analog to Digital Converter.   These circuits are packaged 16 to a card, and the digital data is collected and transmitted by optical fiber to a Trigger Electronics Module (TEM) which is located some 20~m away from the FEE system and services the complete TPC and the Veto System.   Thus each channel in the system transmits its digitized value to the TEM at 1~MS/s.    The TEM synchronizes the entire system, buffers data for 2048~$\mu$s, and forms several varieties of triggers. Upon a trigger, data from 1024~$\mu$s before the trigger and 1024~$\mu$s after the trigger are transferred to a buffer. Subsequently a Data Acquisition (DAQ) PC reads this data and transfers it to storage and to a data sampling system.   The pattern of veto hits is also sampled at 1~MS/s and received by the TEM.

The choice of this $\sim$2~ms ``frame'' is driven by the expected maximum drift time for an electron in the TPC ($\sim$100~$\mu$s), as well as the desire to efficiently capture common correlated backgrounds, such as the delayed coincidence in the $^{214}\textnormal{Bi} \xrightarrow{\beta}$ $^{214}\textnormal{Po} \xrightarrow{\alpha}$ $^{210}\textnormal{Pb}$ decay sequence, where the $^{214}$Po half-life is 164~$\mu$s.  At the 1~MS/s sampling rate, the associated data size is easily manageable.

\subsection{Biasing}
\subsubsection{TPC cathode}
\begin{figure}
	\centering
	\includegraphics[width=6.0in]{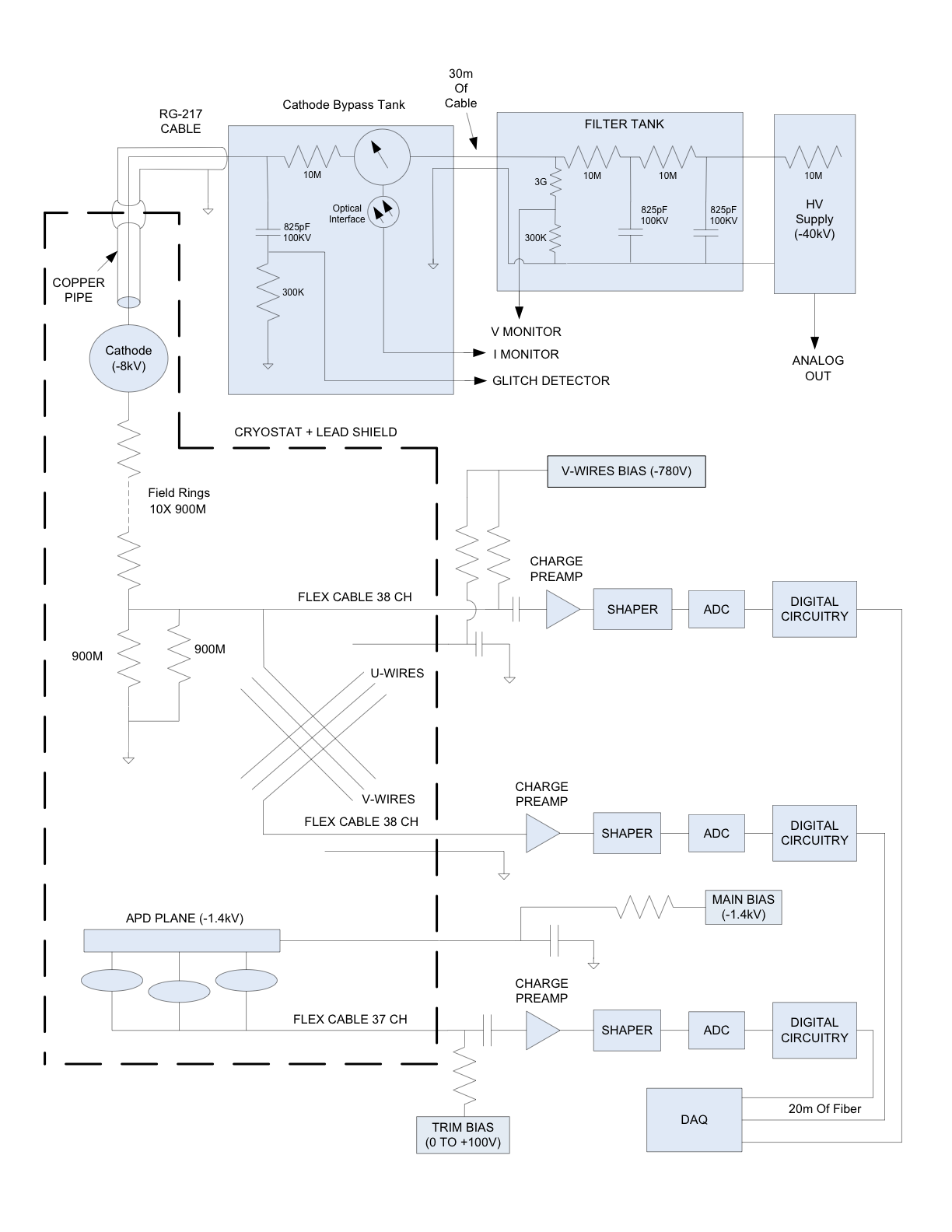}
	\caption{TPC biasing and readout electronics systems. The bias voltages are indicated for a cathode potential of -8~kV.
	\label{fig:electronics}}
\end{figure}

The TPC biasing scheme is illustrated in Figure~\ref{fig:electronics}. While the HV system was nominally designed to operate up to -70~kV, at the time of writing a commercial -40~kV power supply is used to bias the cathode.   The power supply is filtered by two sets of RC filters and connected to the cathode with the custom feedthrough described above.   Both filters are insulated by immersion in a special dielectric fluid~\cite{FC87}.   The first filter is located near the power supply and is separated from the second filter, mounted immediately outside of the front lead shield, by $\sim 30$~m of cable \cite{dielectricsci}.  The second filter includes a floating, battery powered, nano-ammeter shipping out an 18~bit digital signal by optical fiber. This filter also includes a capacitively coupled connection allowing for the observation of small disturbances on the cathode.   The -40~kV power supply is manually set and read out.  The HV system is interlocked to the LXe temperature, pressure, and level in order to prevent accidental damage to the detector. The two filters, including the cables that contribute to the overall parallel capacitance, reduce the power supply noise in the 1 to 100~kHz band by $>110$~dB.

\subsubsection{V-wires and LAAPDs}

As illustrated in Figure~\ref{fig:electronics} the two cathode voltage dividers are terminated at the potential of the V wires that, in turn, are biased by DC-DC converters controlled by the DAQ system.   The LAAPD biasing scheme is also shown in Figure~\ref{fig:electronics}. Both the platter bias and the trim biases are generated by DC-DC converters controlled by the DAQ system.   All bias voltages are read out and interlocked to the LXe temperature, pressure and level.

\subsection{Front end electronics system}
In designing the front-end section of the signal processing, low noise and optimal signal extraction were the main considerations.    In addition, consideration was given to appropriate filtering in order to mitigate possible microphonic effects from vibrations of the polyimide cables biased at high voltage.

The charge preamplifier was based on the design of the discrete component prototype for the BaBar calorimeter~\cite{BaBar_preamp}.  Since only about 300 channels are required, and in order to maintain maximum flexibility in the choice of component values, it was decided to use discrete components rather than develop an ASIC for this application.   The preamplifier is a single-ended folded cascode with a JFET at the input, with an open loop gain for the preamplifier of $\sim$100,000. Large open loop gain is important to ensure an efficient transfer of charge from the capacitive signal source to the feedback capacitor.  For the wire channels the capacitance of 60 to 80~pF is dominated by the polyimide cables, while the 1000~pF capacitance of the LAAPD channels (7 LAAPDs in each channel) is dominated by the devices themselves.   The charge collection efficiency is $> 90$\% for LAAPD channels and 98\% for wire channels.   For LAAPD channels this is achieved using four JFETs in parallel and a 5~pF feedback capacitance, resulting in an effective input capacitance of 50~nF.   Single JFETs are used for the wire channels.  The JFET used~\cite{JFET} has a transconductance of 30~mS at 5~mA drain current.  Using 6~$\mu$s RC-CR shaping the noise was measured to be 2~e$^-$/pF (330~$\mu$V$_\textnormal{rms}$) for LAAPD channels and 1~e$^-$/pF (160~$\mu$V$_\textnormal{rms}$) for wire channels.    These noise figures are close to expectations derived from the transconductance of the input FET.

Bandwidth limiting and extra gain are provided by two operational amplifier stages and associated components, providing differentiation and integration.  At the beginning of EXO-200 data taking in the spring of 2011 the time constants for each of the two stages were set to 3~$\mu$s integration and 10$\mu$s differentiation on all channels.  The initial runs of the detector, showing limited problems from microphonics, led to the decision to increase the system bandwidth for the U (charge collection) wires to 1.5~$\mu$s integration and 40$\mu$s differentiation for the low background run that started in October 2011.  The larger bandwidth allows for better pulse shape discrimination.

ADCs mounted on the same board as the FEE digitize all channels at 1~MS/s and transfer the data to the TEM via optical links.   The FEE is housed in two custom shielded chassis, one for each end of the TPC.    Each chassis includes 3 cards each for U wires, V wires, and LAAPDs, and each card holds 16 channels.   In addition, the chassis contains the DC-DC converters used for the biasing.

\subsection{TEM}
The TEM receives data from the 18 FEE cards and stores them in one of 8 available circular buffers while passing them to a trigger detection circuit. Upon detecting a trigger the current circular buffer is allowed to post-fill for a defined number of samples before being frozen while switching the data stream to the next available buffer. A typical trigger will result in 1024 samples before and after the trigger being transferred to disk. At 1~MS/s this results in a total of 2~MS of data for each trigger. Back to back triggers are handled by enforcing a minimum post fill time (also known as dead time) after each trigger. The trigger logic contained in the TEM is divided into three partitions (one for each element type in the detector) with two groups of trigger levels per partition. The first group of 4 trigger levels receives each channel individually searching for the highest amplitude within a partition (after baseline subtraction). The second group of 4 trigger levels looks at the sum of all of the channels within a partition (channels can be individually disabled) and subtracts a running baseline computed from a configurable number of previous sum results. 

\subsection{DAQ}
The data acquisition (DAQ) is interfaced to the TEM using two PCI-based National Instruments digital I/O cards.  Each card drives a (configurable) 32-bit wide bus running at 20~MHz.  The first I/O card is used for command/control of the front end registers in both the TEM and the front end cards.  It is configured as two 16~bit wide buses, the first to send commands and the second to receive responses.  The second I/O card is configured as a simple 32~bit wide bus to receive data from the TEM.  Both cards are simple I/O streaming devices, so all data received from the TEM must first be software reassembled into identifiable pieces (an event or a command response), then verified for integrity (parity and framing bits). A subset of events is sent in real time to an online monitoring system where it is analyzed for data quality.  Acquired data are written to local direct-attach RAID array storage.   From there they are copied to removable disks outside of the WIPP underground.

Because of the limited bandwidth available at the WIPP site, as these disks fill up, they are unmounted and physically shipped to the offline data storage facility where they are remounted and the data copied once again to secure storage.   The data acquisition is controlled from a graphical web based application which also allows the data quality plots and information from the DAQ database to be viewed at WIPP or remotely by authorized users.

\acknowledgments{EXO-200 is supported by DoE and NSF in the United States, NSERC in Canada, SNF in Switzerland and RFBR in Russia. We thank Applied Plastics Technology, Inc., E. I. du Pont de Nemours and Company, International Rectifier, Flexible Circuit Technologies, Inc., Carriaga Machine, Sheedy Drayage Co., Applied Fusion Inc., Aurubis, JL Goslar, and Advanced Photonix, Inc. In addition, we acknowledge the SNO Collaboration for providing low background acrylic, the KARMEN Collaboration for supplying the veto counters, and WIPP for the hospitality.}

\bibliographystyle{h-physrev3}
\bibliography{exo200ptI-v11}

\end{document}